\documentclass[5p,authoryear]{elsarticle}

\makeatletter
\def\ps@pprintTitle{%
 \let\@oddhead\@empty
 \let\@evenhead\@empty
 \def\@oddfoot{}%
 \let\@evenfoot\@oddfoot}
\makeatother

\usepackage{amsmath}
\usepackage{amsfonts}
\usepackage{amssymb}
\usepackage{graphicx}
\usepackage{xspace}
\usepackage{natbib}
\usepackage{aas_macros}
\usepackage{hyperref}
\usepackage{array}
\usepackage{multirow}
\usepackage{placeins}

\DeclareFixedFont{\ttb}{T1}{txtt}{bx}{n}{8} 
\DeclareFixedFont{\ttm}{T1}{txtt}{m}{n}{8}  

\usepackage{color}
\definecolor{deepblue}{rgb}{0,0,0.5}
\definecolor{deepred}{rgb}{0.6,0,0}
\definecolor{deepgreen}{rgb}{0,0.5,0}
\definecolor{grey}{rgb}{0.4,0.4,0.4}

\usepackage{listings}

\newcommand\pythonstyle{\lstset{
language=Python,
basicstyle=\ttm,
otherkeywords={self},             
keywordstyle=\ttb\color{deepblue},
emph={MyClass,__init__},          
emphstyle=\ttb\color{deepred},    
stringstyle=\color{deepgreen},
frame=tb,                         
showstringspaces=false            %
}}

\newcommand\pythonexternal[2][]{{
\pythonstyle
\lstinputlisting[#1]{#2}}}

\newcommand\sqlstyle{\lstset{
language=SQL,
showspaces=false,
basicstyle=\ttfamily,
numbers=left,
numberstyle=\tiny,
commentstyle=\color{grey},
frame=single
}}

\newcommand{\eagle}{{\sc eagle}\xspace}
\newcommand{\gadget}{{\sc gadget}\xspace}
\newcommand{\subfind}{{\sc subfind}\xspace}
\newcommand{\anarchy}{{\sc anarchy}\xspace}
\newcommand{\dtrees}{{\sc D-Trees}\xspace}
\newcommand{\squotes}[1]{\lq {#1}\rq\xspace}
\newcommand{\dquotes}[1]{\lq\lq {#1}\rq\rq\xspace}
\newcommand{\Msol}{\, \mathrm{M}_\odot}
\newcommand{\HI}{${\rm H~I}$\xspace}
\newcommand{\CIV}{${\rm C~IV}$\xspace}
\newcommand{\OVI}{${\rm O~VI}$\xspace}
\newcommand{\HaloID}{\GalaxyID}  
\newcommand{\GalaxyID}{{\tt GalaxyID}\xspace}
\newcommand{\TopLeafID}{{\tt TopLeafID}\xspace}
\newcommand{\LastProgID}{{\tt LastProgID}\xspace}
\newcommand{\DescendantID}{{\tt DescendantID}\xspace}
\newcommand{\sql}{{\sc sql}\xspace}
\newcommand{\dbAddress}{\url{http://www.eaglesim.org/database.php}}

\begin{document}
\begin{frontmatter}

\title{The \eagle simulations of galaxy formation: public release
  of halo and galaxy catalogues\tnoteref{t1}}

\author[durham]{Stuart McAlpine}
\ead{s.r.mcalpine@durham.ac.uk}
\author[durham]{John C. Helly}
\ead{j.c.helly@durham.ac.uk}
\author[durham]{Matthieu Schaller}
\ead{matthieu.schaller@durham.ac.uk}
\author[durham]{James W. Trayford}
\author[durham]{Yan Qu}
\author[durham]{Michelle Furlong}
\author[durham]{Richard G. Bower}
\author[liverpool]{Robert A. Crain}
\author[leiden]{Joop Schaye}
\author[durham]{Tom Theuns}
\author[tenerife1,tenerife2]{Claudio Dalla Vecchia}
\author[durham]{Carlos S. Frenk}
\author[liverpool]{Ian G. McCarthy}
\author[durham]{Adrian Jenkins}
\author[chile]{Yetli Rosas-Guevara}
\author[mpa]{Simon D. M. White}
\author[gent]{Maarten Baes}
\author[gent]{Peter Camps}
\author[jhu]{Gerard Lemson}

\address[durham]{Institute for Computational Cosmology, Department of Physics,
  University of Durham, South Road, Durham, DH1 3LE, UK}

\address[liverpool]{Astrophysics Research Institute, Liverpool John Moores
  University, 146 Brownlow Hill, Liverpool L3 5RF, UK}

\address[leiden]{Leiden Observatory, Leiden University, P.O. Box 9513, 2300 RA
  Leiden, the Netherlands}

\address[tenerife1]{Instituto de Astrof\'isica de Canarias, C/ V\'ia L\'actea
  s/n, 38205 La Laguna, Tenerife, Spain}

\address[tenerife2]{Departamento de Astrof\'isica, Universidad de La Laguna,
  Av. del Astrof\'isico Francisco S\'anchez s/n, 38206 La Laguna, Tenerife,
  Spain}

\address[chile]{Departamento de Astronom\'ia, Universidad de Chile, Casilla
  36-D, Las Condes, Santiago, Chile}

\address[mpa]{Max-Planck-Institut fur Astrophysik, Karl-Schwarzschild-Str. 1,
  D-85748 Garching, Germany}

\address[gent]{Sterrenkundig Observatorium, Universiteit Gent,
  Krijgslaan 281, B-9000 Gent, Belgium}

\address[jhu]{Department of Physics \& Astronomy, Johns Hopkins
  University, Baltimore, MD, 21218, USA}

\tnotetext[t1]{\dbAddress}

\begin{abstract}
We present the public data release of halo and galaxy catalogues extracted from
the \eagle suite of cosmological hydrodynamical simulations of galaxy
formation. These simulations were performed with an enhanced version of the
\gadget code that includes a modified hydrodynamics solver, time-step limiter
and subgrid treatments of baryonic physics, such as stellar mass loss,
element-by-element radiative cooling, star formation and feedback from star
formation and black hole accretion. The simulation suite includes runs performed
in volumes ranging from 25 to 100 comoving megaparsecs per side, with numerical
resolution chosen to marginally resolve the Jeans mass of the gas at the star
formation threshold. The free parameters of the subgrid models for feedback are
calibrated to the redshift $z=0$ galaxy stellar mass function, galaxy sizes and
black hole mass - stellar mass relation. The simulations have been shown to
match a wide range of observations for present-day and higher-redshift galaxies.
The raw particle data have been used to link galaxies across redshifts by
creating merger trees. The indexing of the tree produces a simple way to connect
a galaxy at one redshift to its progenitors at higher redshift and to identify
its descendants at lower redshift. In this paper we present a relational
database which we are making available for general use. A large number of
properties of haloes and galaxies and their merger trees are stored in the
database, including stellar masses, star formation rates, metallicities,
photometric measurements and mock $gri$ images. Complex queries can be created
to explore the evolution of more than $10^5$ galaxies, examples of which are
provided in appendix. The relatively good and broad agreement of the simulations
with a wide range of observational datasets makes the database an ideal resource
for the analysis of model galaxies through time, and for connecting and
interpreting observational datasets.
\end{abstract}


\end{frontmatter}

\section{Introduction}

Galaxy formation is a complex, non-linear process that involves a wide range of
physical and astrophysical phenomena, from the evolution of dark matter
clustering to intricate feedback effects coupling gas cooling and outflows to
star and black hole formation. Theoretical studies of galaxy formation thus
require rigorous detailed modelling to link together these phenomena over a very
wide range of scales. Two techniques have been developed for this purpose:
semianalytic modelling \citep{White1991} and hydrodynamical simulations
\citep{Carlberg1990,Katz1992}. Both techniques have been extensively developed
over the past 25 years \citep[e.g.][for semi-analytic
  models]{Porter2014,Henriques2015,Lacey2015} and \citep[e.g.][for hydrodynamical
  simulations]{Oppenheimer2010,Puchwein2013,
  Dubois2014,Okamoto2014,Vogelsberger2014,Khandai2015}.

Recently, the Virgo\footnote{\url{http://virgo.dur.ac.uk/}} Consortium's
``Evolution and Assembly of GaLaxies and their Environments'' simulation suite
\citep[\eagle,][]{Schaye2015,Crain2015} has been able to reproduce key
observational datasets, such as the present-day stellar mass function of
galaxies, the correlation of black hole mass and stellar mass and the dependence
of galaxy sizes on stellar mass, with unprecedented fidelity. As well as
reproducing these observations, which were used during the calibration of the
simulation parameters, the simulation outputs match many other properties of the
observed galaxy population and the intergalactic medium both at the present day
and at earlier epochs, as we briefly discuss below. These simulations therefore
provide a powerful resource for understanding the formation of galaxies and for
linking and interpreting observational datasets.

The aim of this paper is to introduce and make available a relational database
that can be queried using the Structured Query Language (\sql) to explore and
exploit the halo and galaxy catalogues of the main \eagle simulations. Columns
containing integrated quantities describing the galaxies, such as stellar mass,
star formation rates, metallicities and luminosities, are provided for more than
$10^5$ simulated galaxies and these can be individually followed through their
evolution across cosmic time. This database is available at the address
\dbAddress.

The simulations follow the gravitational hydrodynamical equations, tracking the
evolution of baryons and dark matter. The initial conditions reflect the small
density fluctuations observed in the cosmic microwave background (CMB). By
tracking the movement of baryon and dark matter particles, the simulations
calculate how these fluctuations are amplified by gravity, and how pressure and
radiative cooling of baryons separate these two matter components of the
universe. The simulations include subgrid formulations to account for processes
that cannot be directly resolved in the calculation and that describe how stars
and black holes form and impact the matter distribution around them. \eagle
improves on previous hydrodynamical simulations of representative volumes,
through the use of physically motivated subgrid source and sink terms as well as
through the adoption of a clear strategy for the calibration of uncertain
subgrid parameters \citep{Crain2015} and by producing a galaxy population that
reproduces many of the characteristics of the observed population over a wide
range of redshifts.

~\\

The usability of the simulation data products is greatly enhanced when presented
in a relational database, making it simple and quick to select galaxy samples
based on multiple galaxy properties, to connect them to their halos and to
follow their evolution over cosmic time \citep{Lemson2006a}.  Such databases
were originally designed to host results from large surveys \citep[e.g. the SDSS
  SkyServer][]{Szalay2000} and later the halo catalogues from dark matter
simulations and galaxy catalogues from semi-analytic models \citep[applied to
  the \emph{Millennium Simulation}, see][]{Lemson2006b}. They have since been
expanded to include the wider range of data available from hydrodynamical
simulations \citep[e.g.][]{Dolag2009,Khandai2015,Nelson2015}. The database
allows multiple indexing of the data that significantly enhances access speed
and allows the selection of smaller data subsets that can be quickly analysed
using simple scripting languages. This approach avoids the need for the user to
copy the raw simulation data or even just the full galaxy catalogues, reducing
data transfer volumes to a manageable level. The galaxy properties stored in the
database can be compared to observations or to other models, whilst the physics
of galaxy formation can be explored by tracking an individual galaxy's behaviour
and environment through cosmic time.

This paper is intended as a reference guide for accessing the publicly available
\eagle database, and is laid out as follows. Section \ref{eagle_sim_suite}
presents a brief overview of the \eagle simulation suite, including the list of
simulations available in the database and the values of the subgrid parameters
that vary, as well as an overview of the construction of the merger trees and
database tables.  A short tutorial describing how to access the data is
presented in Section \ref{database_usage}. We give some words of caution and
some remarks on the simulations in Section~\ref{caveats} and conclude in
Section~\ref{conclusion}. Some additional examples combining the {\sc python}
and \sql languages to access the data are given in \ref{example_scripts} whilst
the full list of galaxy and halo properties available in this data release is
given in \ref{appendix_quantities} together with a list of output redshifts in
\ref{appendix_snapshot} and detailed equations given in \ref{appendix_cosmo}.
Throughout this paper we quote magnitudes in the AB system and use
\squotes{h-free} units unless stated otherwise.

\section{The EAGLE simulation suite}
\label{eagle_sim_suite}

\begin{table*}
\small
\begin{center}
\renewcommand{\arraystretch}{1.5}
\begin{tabular}{lrrrrrrrrrr}
\hline
Identifier & L & N & $m_{\text{g}}$ & $m_{\text{dm}}$ & $\epsilon_{\text{com}}$ & $\epsilon_{\text{phys}}$ & $n_{\text{H,0}}$ & $n_{\text{n}}$ & $C_{\text{visc}}$ & $\Delta \text{T}_{\text{AGN}}$\\
& [cMpc] &  & [M$_{\odot}$] & [M$_{\odot}$] & [ckpc] & [pkpc] & [cm$^{-3}$] & & & [K]\\
\hline\hline
Ref-L0025N0376 & 25 & $2{\times}376^{3}$ & $1.81{\times}10^{6}$ & $9.70{\times}10^{6}$ & 2.66 & 0.70 & 0.67 & 2/ln10 & 2$\pi$ & $10^{8.5}$\\
Ref-L0025N0752 & 25 & $2{\times}752^{3}$ & $2.26{\times}10^{5}$ & $1.21{\times}10^{6}$ & 1.33 & 0.35 & 0.67 & 2/ln10 & 2$\pi$ & $10^{8.5}$\\
Recal-L0025N0752 & 25 & $2{\times}752^{3}$ & $2.26{\times}10^{5}$ & $1.21{\times}10^{6}$ & 1.33 & 0.35 & 0.25 & 1/ln10& 2$\pi{\times} 10^3$ & $10^{9.0}$\\
Ref-L0050N0752 & 50 & $2{\times}752^{3}$ & $1.81{\times}10^{6}$ & $9.70{\times}10^{6}$ & 2.66 & 0.70 & 0.67 & 2/ln10& 2$\pi$ & $10^{8.5}$\\
AGNdT9-L0050N0752 & 50 & $2{\times}752^{3}$ & $1.81{\times}10^{6}$ & $9.70{\times}10^{6}$ & 2.66 & 0.70 &0.67& 2/ln10& 2$\pi{\times} 10^2$ & $10^{9.0}$\\
Ref-L0100N1504 & 100 & $2{\times}1504^{3}$ & $1.81{\times}10^{6}$ & $9.70{\times}10^{6}$ & 2.66 & 0.70 & 0.67 & 2/ln10& 2$\pi$ & $10^{8.5}$\\
\hline
\end{tabular}
\end{center}
\caption{Parameters describing the available simulations. From
  left-to-right the columns show: simulation name suffix; comoving box size;
  total number of particles; initial baryonic particle mass; dark matter
  particle mass; comoving Plummer-equivalent gravitational softening length;
  maximum physical softening length and the subgrid model parameters that vary:
  $n_{\text{H,0}}$, $n_{\text{n}}$, $C_{\text{visc}}$ and $\Delta
  \text{T}_{\text{AGN}}$ (see section 4 of \citet{Schaye2015} for an explanation of their
  meaning).}
\label{table_simulation_list}
\end{table*}

The \eagle simulation suite is a set of cosmological hydrodynamical simulations
in cubic, periodic volumes ranging from 25 to 100 comoving megaparsecs (cMpc) per side that
track the evolution of both baryonic (gas, stars and massive black holes) and
non-baryonic (dark matter) elements from a starting redshift of $z = 127$ to the
present day. All simulations adopt a flat $\Lambda$CDM cosmology with parameters
taken from the \emph{Planck} mission \citep{Planck2013} results: $\Omega_\Lambda
= 0.693$, $\Omega_m = 0.307$, $\Omega_b = 0.04825$, $\sigma_8 = 0.8288$, $n_s =
0.9611$, $Y=0.248$ and $H_0 = 67.77$ km s$^{-1}$ Mpc$^{-1}$ (i.e. $h=0.6777$). The initial
conditions were generated using second-order Lagrangian perturbation theory
\citep{Jenkins2010} and the phase information is taken from the public {\sc
  panphaisa} Gaussian white noise field \citep{Jenkins2013}. Full details of how
the ICs were made are given in Appendix B of \cite{Schaye2015}. The simulation
suite was run with a modified version of the \gadget-3 Smoothed Particle
Hydrodynamics (SPH) code \citep[last described by][]{Springel2005},
and includes a full treatment of gravity and hydrodynamics. The modifications to
the SPH method are collectively referred to as \anarchy (Dalla Vecchia,
(\textit{in prep.}), see also Appendix A of \cite{Schaye2015} and
\cite{Schaller2015b}), and use the ${\cal C}_2$ kernel of \cite{Wendland1995},
the pressure-entropy formulation of SPH of \citet{Hopkins2013}, the time-step
limiters introduced by \cite{Durier2012}, the artificial viscosity switch of
\cite{Cullen2010} and a weak thermal conduction term of the form proposed by
\cite{Price2008}.  The effects of this state-of-the-art formulation of SPH on
the galaxy properties is explored in detail by \cite{Schaller2015b}.

\subsection{Subgrid model}
Processes not resolved by the numerical scheme are implemented as subgrid source
and sink terms in the differential equations. For each process, schemes were
adopted that are as simple as possible and that only depend on the local
hydrodynamic properties. This last requirement differentiates \eagle from most
other cosmological, hydrodynamical simulation projects
\citep[e.g.][]{Oppenheimer2010,Puchwein2013,Vogelsberger2014,Khandai2015} and
ensures that galactic winds develop without pre-determined mass loading factors
and directions, without any direct dependence on halo or dark matter
properties. 

The simulation tracks the time-dependent stellar mass loss due to winds from
massive stars and AGB stars, core collapse supernovae and type Ia supernovae
\citep{Wiersma2009b}. Radiative cooling and heating is implemented
element-by-element following \cite{Wiersma2009a}. Cold dense gas is prevented
from artificial fragmentation by implementing an effective temperature pressure
floor as described by \cite{Schaye_DallaVecchia2008}. Star formation is
implemented stochastically following the pressure-dependent Kennicutt-Schmidt
relation \citep{Schaye_DallaVecchia2008}, with the inclusion of a
metal-dependent star formation threshold designed to track the transition from a
warm, atomic to an unresolved, cold, molecular gas phase, as proposed by
\cite{Schaye2004}. The initial stellar mass function is that given by
\cite{Chabrier2003}. Feedback from star formation is implemented thermally and
stochastically following the method of \cite{DallaVecchia_Schaye2012}. Seed
black holes are placed in haloes greater than a threshold mass of $10^{10}\Msol
/ h$ and tracked following the methodology of \cite{Springel2005a} and
\cite{Booth_Schaye2009}. Gas accretion onto black holes follows a modified
version of the Bondi-Hoyle accretion rate, described by \citep[][, but modified
  as described by \cite{Schaye2015}]{Guevara2013}, and feedback is implemented
following the stochastic AGN heating scheme described by \cite{Schaye2015} and
making use of the energy threshold of \cite{Booth_Schaye2009}. The details of
the implementation and parametrisation of these schemes are motivated and
described in detail by \cite{Schaye2015}.

Because of our limited understanding of these processes and because of the
limited resolution of the simulations, the subgrid source and sink terms involve
free parameters whose values must be determined by comparison of the simulation
results to a subset of the observational data. In the case of \eagle, the
subgrid parameters were calibrated for feedback from star formation and AGN by
using three properties of galaxies at redshift $z=0$, specifically the galaxy
stellar mass function, the galaxy size -- stellar mass relation, and the black
hole mass -- stellar mass relation. The calibration strategy is described in
detail by \citet{Crain2015} who also presented additional simulations to
demonstrate the effect of parameter variations.

Once the simulations have been calibrated using a subset of the observational
data, they can be validated by comparison to additional datasets.  Studies have
so far shown that the simulations broadly reproduce a variety of other
observables such as the $z=0$ Tully-Fisher relation, specific star formation
rates and the column density distribution of intergalactic \CIV and \OVI
\citep{Schaye2015}, the \HI and $\rm{H}_2$ properties of galaxies \citep[Bah\'e
  et al. {\textit submitted},][]{Lagos2015}, the column density distribution of
intergalactic metals \citep{Schaye2015}, galaxy rotation curves
\citep{Schaller2015a}, the $z=0$ luminosity function and colour-magnitude
diagram \citep{Trayford2015}, the evolution of the galaxy stellar mass function
\citep{Furlong2015} and the high-redshift \HI column density distribution
\citep{Rahmati2015}.

\subsection{The simulations in the database}
Table~\ref{table_simulation_list} summarises the simulations that have been
incorporated into the database, including the comoving cubic box length,
baryonic and non-baryonic particle masses and gravitational softening
lengths. Together these parameters determine the dynamic range and resolution
that can be achieved by the simulation. The simulation name includes a suffix to
indicate the simulation box length in comoving megaparsecs (e.g. L0100) and the
cube root of the initial number of particles per species
(e.g. N1504). Simulations with the same subgrid model as the primary run
(L0100N1504) are denoted with the prefix \dquotes{Ref-}. As discussed in
\citet{Schaye2015}, the \dquotes{Recal-} higher-resolution simulation uses
values of the subgrid parameters that have been recalibrated following the same
procedure used for the reference simulation to improve the fit to the $z=0$
galaxy stellar mass function, allowing the user to test the \emph{weak}
convergence of the code\footnote{As discussed in Section~\ref{caveats},
  performing convergence tests is strongly encouraged.}. See \cite{Schaye2015}
for definitions and discussion of the concepts of weak and strong
convergence. Note that Recal-L0025N0752 should be compared to the Ref-L0025N0376
calculation to ensure that the same range of halo mass is sampled in both cases,
eliminating differences due to the simulation volume. To a similar end, the
Ref-L0025N0752 model is provided to allow the user to test the \emph{strong}
convergence of the results. This simulation uses all the subgrid parameters of
the reference model but at a higher mass resolution. All the 25~cMpc volumes
share the same large-scale initial fluctuations, so that objects appear in
(approximately) the same spatial locations in all three runs.

Finally, the database also includes the additional simulation AGNdT9-L0050N0752
that uses a higher AGN heating temperature and increased black hole accretion
viscosity parameter, $C_{\text{visc}}$. As discussed by \cite{Schaye2015}, this
results in a better match to the properties of diffuse gas in galaxy group
haloes, but has only a small effect on the properties of galaxies. This
simulation uses the same initial phases as the Ref-L0050N0752 model, allowing
objects to be matched.

\subsection{Halo, subhalo and galaxy identification}

The raw particle data themselves are not required for many comparisons with
observations. In order to reduce the volume of data to be downloaded and
simplify analysis, we process the simulation outputs individually to locate bound
structures which we identify with galaxies and their associated dark matter
haloes. The processing steps are described in detail by \cite{Schaye2015}. In
brief, overdensities of dark matter are identified using the
\dquotes{Friends-of-Friends} (FoF) method \citep{Davis1985} adopting a linking
length of 0.2 times the average inter-particle spacing. Baryonic particles are
then assigned to the same FoF-halo as their closest dark matter
neighbour. Self-bound \dquotes{subhaloes}, which can contain both baryonic and
dark matter, are later identified using the \subfind algorithm
\citep{Springel2001,Dolag2009} using all particle species.

It is important to note that particles are not shared between subhaloes so that
the correspondence between particles and subhaloes is unique. We identify the
baryonic component of each subhalo with a galaxy and will refer to them as such
from now on. Resolved subhaloes, always have a clear central concentration and
there is a clear identification between the galaxies in the simulations and
galaxies that would be identified in observational studies. Note that small
subhaloes, especially at high redshift, may not contain any stars or even gas
but will still be present in the catalogues. A FoF halo may contain several
subhaloes (or sub-groups in the \subfind terminology); we define the subhalo
that contains the particle with the lowest value of the gravitational potential
to be the \emph{central galaxy} while any remaining subhaloes are classified as
\emph{satellite galaxies} (denoted \texttt{SubGroupNumber} $= 0$ and
\texttt{SubGroupNumber} $> 0$ respectively in the database nomenclature, see
below).

The stellar mass assigned to a galaxy may include diffuse particles at a large
distance. Such particles make up the intra-cluster/intra-group light and would
not normally be included in a galaxy's photometry. We therefore also include
aperture-based measurements in the database. 

Exceptionally, \subfind may identify an internal high-density component of a
galaxy as a distinct subhalo. Such spurious identifications are discussed in
Sec.~\ref{caveats} and are labelled in the main database table with the field
{\tt Spurious}.

For each simulation we release 29 snapshot outputs between redshift $20$ and $0$
(the full list of released output redshifts is given in the Appendix table
\ref{table:snapshot_list}). We later analyse the properties of each subhalo in
post-processing in order to calculate galaxy and subhalo properties, such as
stellar masses, galaxy sizes, star formation rates and luminosities. Each
subhalo and hence each galaxy is assigned an index, its \HaloID, that allows one
to identify an object uniquely both in space and time. Note that since the
\HaloID is unique to a particular output redshift, a galaxy will change its
\HaloID over time. The 29 catalogues of galaxies are then linked through time
via a galaxy merger tree, allowing one to track the evolution of a galaxy
(through the evolution of its \HaloID) with time.  The construction and
structure of these trees is presented in Section \ref{subsection:merger_trees}.

\subsection{Integrated quantities}
At each redshift the galaxies are processed one-by-one to produce integrated
quantities from the raw particle information. These are the quantities stored in
the different tables of the database.

For the simplest quantities, such as galaxy mass, metallicity or star formation
rate, the post-processing only involves a simple summation over the particles
but other quantities, such as luminosities in various filters, require much more
involved calculations. The full list of quantities present in the database,
together with a description of the post-processing operations performed, is
given in \ref{appendix_quantities}. To allow for an easier comparison with
observational measurements, masses, star formation rates and velocity
dispersions are also computed within fixed spherical apertures.

\subsection{Mock gri images}

\begin{figure*}
\centering
\includegraphics[width=0.32\textwidth]{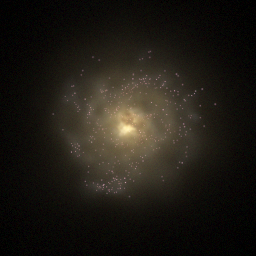}
\includegraphics[width=0.32\textwidth]{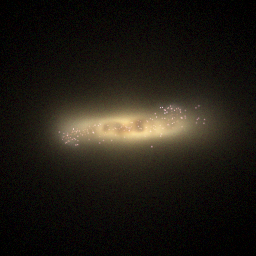}
\includegraphics[width=0.32\textwidth]{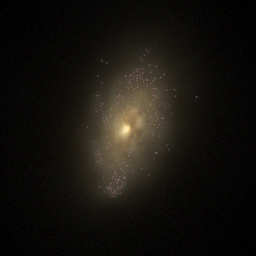}
\caption{Mock $gri$ images of a galaxy at $z=0.1$ as available in the database. The left,
  central and right panels correspond to the {\tt Image\_face}, {\tt
    Image\_edge} and {\tt Image\_box} views (in the database nomenclature) of
  the same simulated galaxy ({\tt GalaxyID}$~=16116800$ in the Ref-L0100N1504
  simulation). The images are 60~pkpc on a side. Note the clear presence of a
  bulge, of dust absorption and of spiral arms.}
\label{fig:mockImage}
\end{figure*}

For visualisation purposes, images are provided for galaxies with 30 physical
kpc (pkpc) aperture stellar masses $> 10^{10} {\rm M}_\odot$. Images are
generated from mock observations made using the {\sc skirt} code \citep{Baes03,
  Camps15}, with \textsc{galaxev} \citep{BC03} and \textsc{mappings iii}
\citep{Groves08} spectra to represent star particles and young H\textsc{ii}
regions respectively, as described by Trayford et.al., (\textit{in prep.}). A
square field of view of 60~pkpc on a side is used for observations in the SDSS
$gri$ bands \citep{SDSSfilters}, with the galaxy spectra red-shifted to $z =
0.1$ to approximate SDSS colours. No artificial seeing is added to the
images. Each galaxy above the stellar mass threshold is observed \textit{face-}
and \textit{edge-on} to the galactic plane, defined using the stellar angular
momentum vector within 30~pkpc. A `\textit{box}' projection is also provided,
with galaxies viewed down the simulation z-axis and the horizontal and vertical
image axes corresponding to the simulation $x$ and $y$ axes respectively. The
three-colour $gri$ images are prepared from the virtual {\sc skirt} observations
adopting the method of \citet{Lupton04}. Figure \ref{fig:mockImage} shows these
three images for the same example galaxy.


\subsection{Merger trees}
\label{subsection:merger_trees}
As galaxies rarely evolve in isolation, they are subject to mergers with
neighbouring galaxies.  This adds serious complexity to tracing the history of
an individual galaxy from the present-day to its formation and as such we must
construct a \emph{merger tree} to connect galaxies across simulation output
times. Descendant subhaloes and hence galaxies are identified using the \dtrees
algorithm \citep{Jiang2014}, with a complete description of its adaptation to
the \eagle simulations provided in Qu et al. (\textit{in prep.}). In essence,
the algorithm traces subhaloes using the $N_{\text{link}}$ most bound particles
of any species, identifying the subhalo that contains the majority of these
particles as a subhalo's descendant at the next output time.  We define $N_{\rm
  link} = \min(100, \max(0.1N_{\rm galaxy}, 10))$, where $N_{\rm galaxy}$ is the
total number of particles in the parent subhalo.  This allows the identification
of descendants, even in the case where most particles have been stripped and it
minimises the misprediction of mergers during fly-bys \citep{Fakhouri08,
  Genel09}.

The galaxy with the most $N_{\rm link}$ particles at the next output is
identified as the single \emph{descendant} of a galaxy, while a descendant
galaxy can have multiple progenitors. The trees are stored in memory following
the method introduced by \cite{Lemson2006a} for the \emph{Millennium Simulation}
\citep[See also the supplementary material of][where the details of the tree
  ordering are summarized]{Springel2005b}. However, the
\emph{main progenitor}, corresponding to the main branch of the tree, is defined
as the progenitor with the largest \squotes{branch mass}, i.e., the mass summed
across all earlier outputs as proposed by \cite{DeLucia_Blaizot2007}. This
definition of the main progenitor, as opposed to the simple definition of the
progenitor with the largest mass, is used to avoid main branch swapping in the
case of similar-mass mergers, as explained by Qu et al. (\textit{in
  prep.}). Note that because the progenitor with the largest branch mass
determines the main branch of the tree, main branch galaxies do not necessarily
correspond to the central galaxy (or \texttt{SubGroupNumber} $= 0$ galaxy) of a
given halo.

There are two further aspects of the merger trees that must be kept in mind when
analysing the simulation:
\begin{itemize}

\item A galaxy can disappear from a snapshot but reappear at a later time
  (e.g. if one galaxy passes through another one).  To account for this,
  descendants are identified using up to 5 snapshots at later times.

\item Care must be taken when determining mass ratios, for example in the case
  of mergers, as galaxies can lose or gain mass due to the definition of the
  subhaloes.
\end{itemize}

Both of these relatively rare cases are considered further by Qu et
al. (\textit{in prep.}), who discuss their impact on the assembly of galaxy
mass.\\

\begin{figure*}[t]
\centering\includegraphics[width=\textwidth]{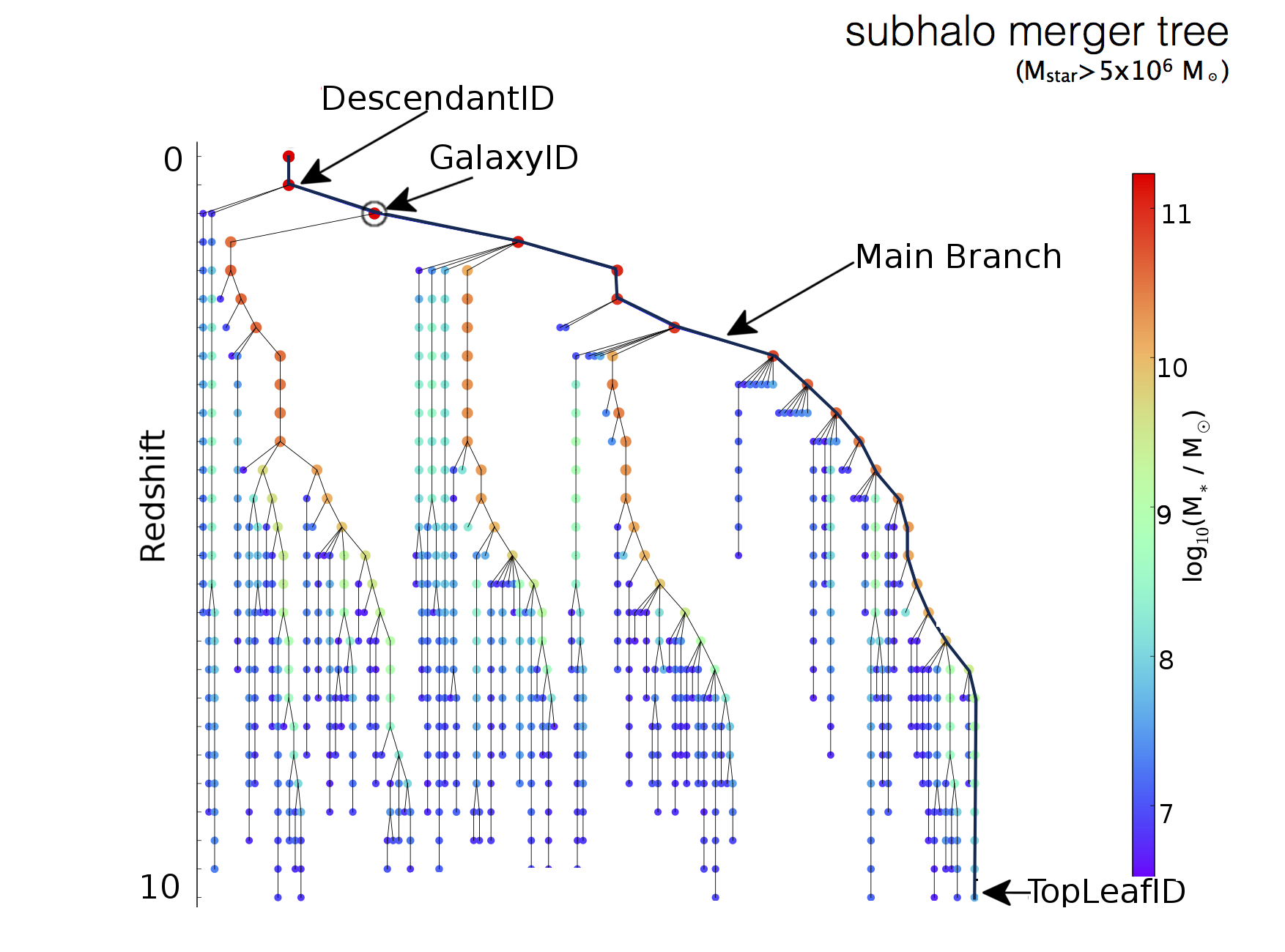}
\caption{Merger history of a galaxy with a $z=0.18$ stellar mass M$_*\sim10^{10}
  \Msol$ indicated by the circled dot. Symbol colours and sizes are scaled with
  the logarithm of the stellar mass. The \HaloID of this galaxy points towards
  it, as indicated by the arrow. The main progenitor branch is indicated with a
  thick black line, all other branches with a thin line. The \TopLeafID gives
  the \HaloID of the highest redshift galaxy on the main progenitor branch
  whilst the \LastProgID (not shown) gives the maximum \HaloID of all the
  progenitors of the galaxy considered. Querying all galaxies with an ID between
  \GalaxyID and \LastProgID will return all the progenitor galaxies in the
  tree.}
\label{fig:mergerTree}
\end{figure*}


\subsection{Technical aspects and infrastructure}
\label{subsection:technical_aspects}

Multiple layouts and frameworks are available for storing large datasets (such
as \textsc{MongoDB}\footnote{\url{https://www.mongodb.org/}},
\textsc{SciDB}\footnote{\url{http://www.paradigm4.com/}},
\textsc{Hadoop}\footnote{\url{http://hadoop.apache.org/}}, ...) each coming with
advantages and shortcomings. In the case of galaxy catalogues extracted from
cosmological simulations, the \textit{Millennium simulation} used an \sql
database for its public release and the wider astronomy community has since
developed a familiarity with its structure and way to query the data. To allow
users the simplest transition between databases, we have adopted the same
framework and a similar table design as the \textit{Millennium simulation} \sql
database \citep{Lemson2006b}. More efficient ways of querying the data could
exist, with differing database formats or table structures, however we decided
that maintaining the familiar aspects of previous databases outweighs the
potential performance gains.

The server hosting the front end web interface operates on Centos linux, running
\textsc{Apache Tomcat 6.0.24}. This server interfaces with the database host,
submitting queries and having their results streamed via a Java web application
(originally written for the \textit{Millennium simulation}). The database itself
is stored on a single physical Windows Server 2008 system with 128~GB of ram,
80~TB of disk storage and two Xeon E5-2670 CPUs which runs Microsoft SQL Server
2012. The main table for the largest simulation contains 65,996,151 rows, which
corresponds to $\approx300~\rm{GBytes}$ of disk space.

Columns are indexed on disk as follows (see below for the description of the
content of each table):

\begin{enumerate}

\item The \textbf{SubHalo} and \textbf{Sizes} tables have a clustered index on
  the \GalaxyID. This allows joins between the tables and queries for
  progenitors and descendants to run efficiently. \GalaxyID rows are assigned
  such that progenitors of each galaxy have a continuous range of \GalaxyID.

\item The \textbf{SubHalo} tables have an additional index on (\texttt{Snapnum},
  \GalaxyID) due to the common nature of queries that request a particular time
  in the simulation.

\item The \textbf{Aperture} tables have a clustered index on the combination of
  (\GalaxyID, \texttt{ApertureSize}) and \\(\texttt{ApertureSize}, \GalaxyID) to
  aid queries searching for all information about a single galaxy or one
  aperture size for many galaxies respectively.

\item The \textbf{FOF} tables are clustered on the combination
  (\texttt{SnapNum}, \texttt{GroupID}), which uniquely identifies the FoF group
  and can be used to join to the \textbf{SubHalo} table.

\end{enumerate}

Typical queries (such as the ones given as examples in
Sec.~\ref{database_usage}) take a few milliseconds to complete on the
server.  More complex queries (i.e. joining multiple tables or
navigating the merger trees for multiple galaxies at the same time)
can take up to a few seconds. As the usage goes up, additional
indexing of the columns could be added to improve the performance of
common, more complex, queries.

The mock \emph{gri} images have been processed once for the entire simulation
and are stored on a separate server. When querying images, the \sql server
generates valid HTML tags containing the links to the images. No caching has
been put in place but such facility could easily be added in case of large
demand.

%
\section{Use of the database}
\label{database_usage}

This section provides an overview of the database interface and of the different
tables available for each simulation. Simple examples of how to query and
combine the tables are presented.

\subsection{Database interface}

\begin{figure*}  [p]
\centering\includegraphics[width=0.98\textwidth]{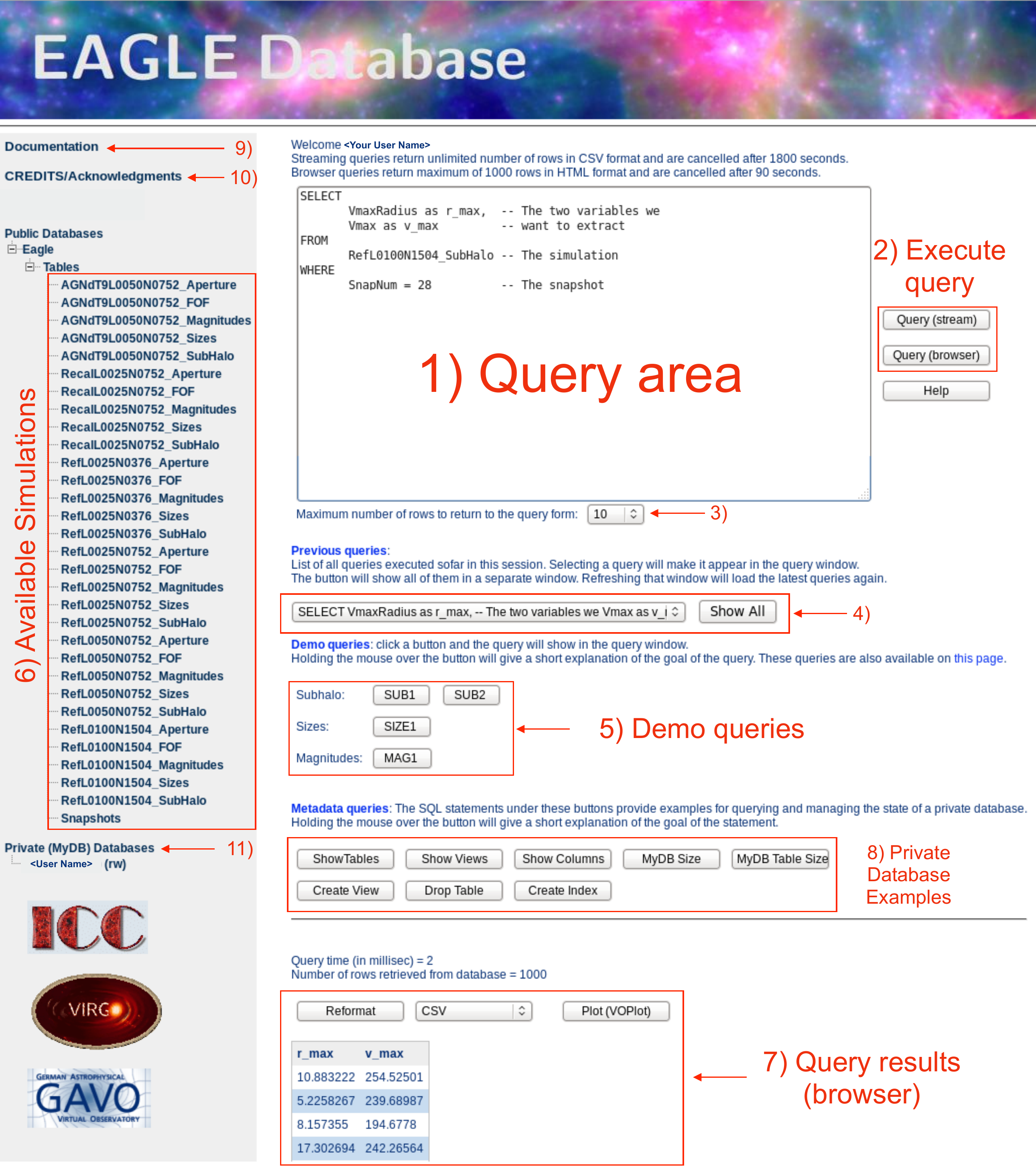}
\caption{The interface of the \eagle database. \sql queries should be entered
  into the query area (1) and can be executed either via the \squotes{browser}
  or \squotes{stream} buttons (2). The \emph{browser query} returns a limited
  number of results (3) at the bottom of the page (7), pressing the
  \emph{Reformat} button will then return the full results in the selected
  format (default CSV) and \emph{Plot(VOPlot)} is a simple way to visualise the
  data. This is the easiest method to test \sql scripts. The \emph{stream query}
  returns all the results in a CSV format in a separate window to ease their
  download to a local device. Previous queries can be restored from the
  drop-down menu (4). The example query buttons (5) insert example \sql
  queries into the query area to help new users with the syntax and structure of
  the database. Similarly, examples creating and managing a private database are
  generated by clicking on the buttons (8). The list of available simulations
  and tables is given on the left hand side (6) with links to further
  documentation describing their contents.  Users' own database tables are
  listed below (11). Further step-by-step documentation on how to use the web
  interface is provided (9) as well as links pointing to credits and
  acknowledgements (10). }
\label{fig:interface_example}
\end{figure*}

The main interface to the \eagle database is shown in
Figure~\ref{fig:interface_example}. Users familiar with the {\it Millennium}
database \citep{Lemson2006b} and its clones will recognize the main features of
the interface and should be able to adapt their scripts easily to the \eagle database.

\sql queries can be typed in the main text box (number 1 in the Figure) and are
submitted to the database by pressing either of the buttons to the right (number
2). Some help with \sql queries can be obtained by clicking on the corresponding
button. The results of queries submitted to the \emph{browser} are returned at
the bottom of the page in the form of an HTML table\footnote{Note that the
  \emph{browser} queries time out after $90$ seconds. More substantial queries
  should be submitted via the \emph{stream} queries option. These only time out
  after $30$ minutes.} (number 7). This allows users to submit small queries and quickly
verify the syntax. If images are being queried, they will appear directly in the
results table. Larger, more complex queries should be submitted to the
\emph{stream} and will be returned in Comma-Separated-Value (CSV) format in a
new window. The number of rows returned by the \emph{browser} queries can be
specified via the drop-down menu (number 3). The \emph{stream} queries always
return all rows. Previous queries can be recovered using the drop-down
menu (number 4).

The queries from this paper are available as examples (number 5). These can
later be adapted to match the user's need. All the available simulations and
their tables are listed in the left-hand panel (number 6) with links to the
documentation describing each entry in the table. All registered users receive a
private database (MyDB) in which they can store query results for further
processing at a later date. A link to MyDB is provided (number 11). Examples of
how to create and manage such private tables can be obtained by clicking on the
buttons at the bottom of the screen (number 8). Finally, some documentation, a
list of credits are given at the top of the page (numbers 9 \& 10).

\subsection{Galaxy merger-tree traversal}
In order to simplify the navigation of the trees, the database is stored with
depth-first ordering \citep[see][Qu et al., \textit{in prep.}]{Lemson2006a}. The
progenitors of a galaxy can then easily be identified. To allow simple
traversing of the merger tree of a given galaxy (with its unique \HaloID), three
additional columns are assigned to each galaxy:
\begin{enumerate}
\item \TopLeafID: This is the \HaloID of the highest-redshift main branch
  progenitor. \\ 
  All the galaxies on the \emph{main progenitor branch} of a galaxy with \HaloID $i$
  and \TopLeafID $j$ have a \HaloID in the range $[i,j]$ in ascending redshift order.
\item \LastProgID: This is the maximum \HaloID of all progenitors irrespective
  of their branch. \\
  All the galaxies on \emph{any progenitor branch} of a galaxy with \HaloID $i$
  and \LastProgID $k$ have a \HaloID in the range $[i,k]$. 
\item \DescendantID: This is the \HaloID of the unique descendant galaxy of
  $i$. \\
  If no descendant galaxy is identified then the \\ \DescendantID of a galaxy is
  set to its own \HaloID.
\end{enumerate}
In Fig. \ref{fig:mergerTree} we show a merger tree for a typical galaxy,
indicated by its \HaloID. The main branch is shown using a thicker blue line
and the IDs required to navigate the tree are shown with arrows pointing
towards the galaxy to which they correspond in the tree\footnote{Users
  familiar with the \emph{Millennium} database can modify they queries
  by replacing {\tt HaloID} with \GalaxyID, {\tt mainLeafID} or {\tt
    endMainBranchID} with \TopLeafID and {\tt lastProgenitorId}
  with \LastProgID. Note also that in the \emph{Millennium} database, a
galaxy with no descendant has its \DescendantID set to -1 and not to
\GalaxyID as in the \eagle database.} . \\ Examples using the
\sql language showing how to traverse the tree forwards and backwards in time
are provided in \ref{example_scripts}.

\subsection{Content of the database}
The \eagle database for each simulation has information distributed
across five \sql \lq tables\rq\ listed in Table~\ref{table_tables}, whose
contents are detailed in \ref{appendix_quantities}. \\

\begin{table}
\footnotesize
\begin{center}
\renewcommand{\arraystretch}{1.5}
\begin{tabular}{ll}
\hline
\sql Table name & Contents \\
\hline\hline
{\bf SubHalo} & Main galaxy properties \\
{\bf FOF} & Halo properties\\
{\bf Sizes} & Galaxy sizes \\
{\bf Aperture} & Galaxy properties in 3D apertures \\
{\bf Magnitudes} & Galaxy photometry in the GAMA bands\\
\hline
\end{tabular}
\end{center}
\caption{\sql tables available for each simulation. The tables are prefixed with
  the name of the simulation to which they correspond. For example, the table of
  magnitudes for the 50~Mpc Ref- model is labelled {\bf
    RefL0050N0752\_Magnitudes} as can be seen on Figure
  \ref{fig:interface_example}.}
\label{table_tables}
\end{table}

{\bf SubHalo:} This is the main table containing properties of {\em galaxies},
for example masses (of dark matter, gas, stars, and black holes), star formation
rates, metallicities and angular momentum. The \GalaxyID of a galaxy can be used
to navigate through its descendants and progenitors as well as to join the
galaxy property table to other tables containing additional properties. The
examples below demonstrate how to do this.\\ A full description of the contents
of the {\bf SubHalo} table is given in Table \ref{table:subhalo1}.\\

{\bf Aperture:} This table contains masses, star formation rates and velocity
dispersions measured in a range of spherical apertures. Table \ref{table:aperture}
gives a full list of the fields present in that \sql table. This table can be
joined to the {\bf SubHalo} table via the \GalaxyID of the objects. \\

{\bf Magnitudes:} This table contains non-dust-attenuated rest-frame broad-band
magnitudes in the SDSS $ugriz$ filters \citep{SDSSfilters} and in the UKIRT
$YJHK$ filters \citep{UKIRTfilters}, computed in 30~pkpc spherical apertures for
all galaxies with stellar mass greater than $10^{8.5}\Msol$ as described in
\cite{Trayford2015}.  See Table \ref{table:magnitudes} in the appendix for more
details. This table can be joined to the {\bf SubHalo} table via the \GalaxyID
of the objects. \\

{\bf Sizes:} This table contains half-mass sizes of galaxies computed starting
from apertures, as presented in Furlong et al., ({\it in prep.}). See Table
\ref{table:sizes} in the appendix for a full list of available quantities. This
table can be joined to the {\bf SubHalo} table via the \GalaxyID of the
objects. \\

{\bf FOF:} This table contains properties of {\em haloes}, for example mass and
spherical overdensity radii. A full description of the contents of the FOF group
table, including the units and dimensions of each variable, is given in Table
\ref{table:fof}. This table can be joined to the {\bf SubHalo} table via the
    {\tt GroupID} of the galaxies, given in the {\bf SubHalo} table. \\

The {\bf FOF} and {\bf SubHalo} tables also contain a field with random number
uniformly distributed in the range $[0,1)$ allowing the users to generate
  unbiased sub-samples of galaxies or haloes.

\subsection{Querying the database tables}
In this section we will illustrate the use of the database by
presenting simple example queries showing the basic usage of the different
\sql tables. \\

The queries can be typed directly into the web interface or used in a {\sc
  Python} script, as described in \ref{example_scripts}, or using the UNIX {\tt
  wget} command as described in the online documentation. The first example
illustrates how to query the main galaxy table ({\bf SubHalo}) in order to plot
the relation between $r_{\max}$ and $v_{\rm max}$ at $z=0$ ({\tt Snapnum}$~=28$)
for the Ref-L0100N1504 simulation. In the database nomenclature, these
quantities are {\tt VmaxRadius} and {\tt Vmax} (see Table \ref{table:subhalo2}).

\noindent 
The \sql command to be typed in the input window is
{
\sqlstyle
{
\footnotesize
\begin{lstlisting}[numbers = none,caption={Generate
      $r_{\max}$-$v_{\max}$ table at $z=0$}]
SELECT        
   VmaxRadius as r_max,  -- The two variables we
   Vmax as v_max         -- want to extract
FROM 
   RefL0100N1504_SubHalo -- The simulation
WHERE  
   SnapNum = 28          -- The snapshot 
\end{lstlisting}
}
}
\noindent Clicking on the ``Query (stream)'' will open a new window containing
the resulting two-column table with headers ``r\_max'' and ``v\_max''
in CSV format. \\

For many applications, multiple \sql tables have to be queried at the same
time. The properties of a galaxy can be retrieved across the tables by joining
their \GalaxyID. A rest-frame colour-magnitude diagram using the SDSS $g$ and
$r$ bands at $z=0.1$ ({\tt SnapNum}~$=27$) for central galaxies ({\tt
  SubGroupNumber}$~=0$) with a stellar mass larger than $10^9\Msol$ ({\tt
  Mass\_Star}~$>{\mathtt{1.0e9}}$) in a $30~\rm{pkpc}$ aperture ({\tt
  ApertureSize}~$=30$) can be constructed by joining the {\bf SubHalo} table to
the {\bf Magnitudes} and {\bf Aperture} ones.

This query reads
{
\sqlstyle
{
\footnotesize
\begin{lstlisting}[numbers = none,caption={Generate
      table of $g-r$ vs. $r$ colour-magnitude table for central
      galaxies with $M_*>10^9$ at $z=0.1$}]
-- Select the quantities we want
SELECT        
   (MAG.g_nodust - MAG.r_nodust) as g_minus_r,
   MAG.r_nodust as r
-- Define aliases for the three tables
FROM 
   RefL0100N1504_SubHalo as SH,
   RefL0100N1504_Magnitudes as MAG,
   RefL0100N1504_Aperture as AP
-- Apply the conditions
WHERE  
   SH.SnapNum = 27 and        -- z=0.1
   SH.SubGroupNumber = 0 and  -- Centrals only
   AP.Mass_Star > 1.0e9 and  -- Mass limit
   AP.ApertureSize = 30 and  -- Aperture size
-- Join the objects in the 3 tables
   SH.GalaxyID = MAG.GalaxyID and
   SH.GalaxyID = AP.GalaxyID
\end{lstlisting}
}
}
\noindent and will return a two-column table with ``g\_minus\_r'' and ``r'' as
headers containing the colours and $r$-band magnitudes of the selected
galaxies.

Note that, as discussed in Section~\ref{caveats}, we recommend to always use
quantities measured in apertures to avoid incorporating intra-cluster light into
mass or star formation rate estimates. \\

Another common use of the database is to track one galaxy across time. To this
end, one can navigate through the main progenitor branch. This final example
tracks an interesting object (\GalaxyID$=1848116$) discovered at redshift $z=1$
through time and constructs the stellar metallicity evolution accompanied by
the mock $gri$ face-on images of the object at all redshifts. One hence has to
construct a query that returns all of the descendants (on the main branch) of
the object by finding all galaxies that have the interesting object's \GalaxyID
between their own \GalaxyID and their {\tt TopLeafID}. To get the progenitors
one additionally requests all galaxies with \GalaxyID between the interesting
object's \GalaxyID and its {\tt TopLeafID}.  This demonstrates the merger tree
navigation introduced in Section \ref{subsection:merger_trees}. Note that
adding conditions on the snapshot number ({\tt SnapNum}) helps speed up the
queries dramatically. This query reads

{
\sqlstyle
{
\footnotesize
\begin{lstlisting}[numbers = none,caption={Returns the evolution along the main
branch of stellar metallicity with redshift of a given galaxy with its images.
To return the evolution along all branches replace \tt{TopLeafID} with
\tt{LastProjID} in line 20.}]
-- Select the quantities we want
SELECT      
   SH.Redshift as z,      
   SH.Stars_Metallicity as Z,      
   SH.Image_Face as face      
-- Define two aliases for the main table
FROM      
   -- Properties we want to extract      
   RefL0025N0752_Subhalo as SH,      
   -- Acts as a reference point      
   RefL0025N0752_Subhalo as REF      
-- Apply the conditions      
WHERE      
   REF.GalaxyID=1848116 and -- GalaxyID at z=1      
   -- To find descendants      
   ((SH.SnapNum > REF.SnapNum and REF.GalaxyID      
   between SH.GalaxyID and SH.TopLeafID) or 
   -- To find progenitors  
   (SH.SnapNum <= REF.SnapNum and SH.GalaxyID  
   between REF.GalaxyID and REF.TopLeafID))  
-- Order the output by redshift      
ORDER BY      
   SH.Redshift
\end{lstlisting}
}
}
\noindent and will return a sorted table with a redshift and a metallicity
column as well as a column containing the postage-stamp images of the galaxy at
each redshift when using the ``Query (browser)'' button. These examples along with
the more complex queries are given in \ref{example_scripts} are listed on the
webpage documentation.


\section{Recommendations, caveats and credits}
\label{caveats}

\subsection{Caveats regarding the usage of the data}

In this section we list a series of recommendations and known limitations that the
authors have uncovered while working on the analysis of the simulation and the
preparation of the database. These points should be taken into consideration to
exploit the simulation outputs fully and to avoid mistakes in the interpretation
of the results.

\paragraph{\bf Finite resolution}
When using the galaxy catalogues, it should be remembered that the properties of
low-mass galaxies should be treated with caution. Large numbers of particles are
required to adequately sample the formation history of a galaxy. In general, we
find that many galaxy properties are unreliable below a stellar mass of $10^{9}
\Msol$ for the intermediate resolution simulations \citep{Schaye2015}. For any
given quantity, these effects can be assessed by comparing the Ref-L0025N0376
simulation with the higher-resolution Recal-L0025N0752 and Ref-L0025N0752
simulations.

\paragraph{\bf Finite volume}
Although the main simulation is one of the largest of its kind, its volume is
still only $10^{-3}~\rm{cGpc}^3$, a volume much smaller than the volumes
typically probed by surveys of the extragalactic Universe. This implies that
rare objects are unlikely to be found in the simulation volume. Moreover, due to
missing large-scale modes, the number density of rare objects will typically be
underestimated. Only a handful of haloes with mass $M_{200}$
(\texttt{Group\_M\_Crit200} in the {\bf FOF} table) above $10^{14}\Msol$ are
present in the main simulation, limiting the analysis of cluster-like
objects. The convergence with box size can be assessed by comparing the main
simulation to the smaller volumes that use the same resolution.

\paragraph{\bf Aperture masses and SFRs} 
The stripping of satellite galaxies as they orbit within a halo generates a
significant mass loss at large radii. The resulting diffuse light (and any
diffuse star formation) is extremely difficult to observe and is not commonly
included in observational galaxy catalogues.  Furthermore, the total galaxy
stellar masses and star formation rates can depend strongly on the precise
assignment of particles to the main subhalo within each FOF group by the
\subfind algorithm, which can lead to spurious total mass evolution.  For these
reasons, studies published by the EAGLE team use aperture masses and star
formation rates, typically in an aperture of $30~\rm{pkpc}$. As discussed by
\cite{Schaye2015}, this corresponds roughly to an $R_{80}$ Petrosian aperture
and is hence particularly well-suited to comparison with observations. We
recommend the use of aperture values when available.

\paragraph{\bf Self-bound star clusters and black holes}
As discussed by \cite{Schaye2015}, small dense stellar regions within galaxies
may occasionally be identified by \subfind as distinct subhaloes and hence
'galaxies'. These appear in the catalogue as rather unusual objects with little
stellar mass but anomalously high metallicity or black hole mass. These
``spurious'' galaxies are flagged in the database in the column {\tt Spurious}
(see the table in \ref{appendix_quantities}). Such objects should not be
considered as genuine galaxies and should be discarded from samples of simulated
galaxies.

\paragraph{\bf Black hole masses and accretion rates}
The black hole masses given in the main table (table {\bf SubHalo}, column {\tt
  BlackHoleMass}) do not directly correspond to the mass of the central
supermassive black hole of a galaxy, but to a summed value of all black holes
assigned to that subhalo. For cases where {\tt
  BlackHoleMass} $> 10^6\Msol$ this closely approximates the mass of
the most massive black hole. {\tt MassType\_BH} refers to the sum of the black
hole particle masses (see \ref{appendix_cosmo} for details of particle and
subgrid masses) and therefore should not be used for a galaxy's black hole
mass. Similarly, due to the coarse time sampling of the outputs, the high
temporal variability of the black hole accretion rates cannot be captured in the
database outputs and as such the quantity {\tt BlackHoleMassAccretionRate} should be
treated with great care.

\paragraph{\bf Stellar velocity dispersion and morphology}
The field {\tt StellarVelocityDispersion} stored in the {\bf SubHalo} table
is a measure of the kinetic energy of the stars, $\sigma = \sqrt{2 E_{\rm{K}} /
  3 M}$, and not a measure of the amount of stellar kinetic energy in dispersion
as opposed to rotation. In particular, it cannot be used to distinguish
rotationally supported galaxies (spirals) from dispersion supported galaxies
(ellipticals).

\paragraph{\bf Galaxy images and magnitude tables}
The images provided in the database are generated using only the particles
within a particular subhalo, in order to correspond with an entry in the
database tables. As a result satellites or merging partners may not be visible
in the images. While the images are observed as if redshifted to $z=0.1$ to
approximate typical SDSS colours, the magnitude tables are measured in the
rest-frame. The inclusion of different population synthesis models, dust
absorption and the relative scaling of images also implies that images are not
reducible to magnitude table entries. 

\paragraph{\bf This simulation is not the real Universe} 
The papers presenting \eagle have shown that the simulation broadly reproduces a
wide set of observational properties of galaxies and the intergalactic
medium. When using the database it should nevertheless be remembered that there
are known discrepancies between the simulation results and observational
data. In particular, we highlight the following points:

\begin{itemize}
\item Although the $z=0.1$ stellar mass function was used in the calibration of
  the simulation, the stellar mass density is approximately $20\%$ lower than
  inferred from observations \citep{Schaye2015,Furlong2015}. This missing mass
  can be related to the slight undershoot of the ``knee'' of the simulated
  galaxy stellar mass function.

\item The evolution of specific star formation rates broadly follows the trends
  seen in observational data, but with a normalisation lower by, depending on
  redshift, 0.3 - 0.5 dex \citep{Schaye2015, Furlong2015}. Note, however, that
  the \eagle galaxies are in good agreement with the recent recalibration of
  star formation indicators by \cite{Chang2015} \citep[see Fig. 5 of ][]{Schaller2015b}.

\item The present-day stellar mass -- metallicity relation in the
  intermediate-resolution Ref- model is flatter than the one inferred from
  observational data \citep{Schaye2015}.  Note, however, that the relation
  becomes steeper in the higher-resolution Recal-L0025N0752 simulation, in
  agreement with the observations.

\item The transition from active to passive galaxies occurs at too high a
  stellar mass at $z=0$ \citep{Schaye2015, Trayford2015}.

\end{itemize}

This list of flaws is certainly not exhaustive. Future papers will
undoubtedly uncover further deficiencies.

\subsection{Acknowledgement of usage}
To recognise the effort of the individuals involved in the design and execution
of these simulations, in their post-processing and in the construction of the
database, we kindly request the following:
\begin{itemize}

\item Publications making use of the \eagle data extracted from the public
  database are kindly requested to cite the original papers introducing the project
  \citep{Schaye2015,Crain2015} as well as this paper (McAlpine et al., 2015).

\item Publications making use of the database should add the following line in
  their acknowledgement section: ``\textit{We acknowledge the Virgo Consortium
    for making their simulation data available. The \eagle simulations were
    performed using the DiRAC-2 facility at Durham, managed by the ICC, and the
    PRACE facility Curie based in France at TGCC, CEA, Bruy\`eres-le-Ch\^atel.}''.

\item Furthermore, publications referring to specific aspects of the subgrid
  models, hydrodynamics solver, or post-processing steps (such as the
  construction of images or photometric quantities, and the construction of
  merger trees), are kindly requested to not only cite the above papers, but
  also the original papers describing these aspects. The appropriate references
  can be found in section 2 of this paper and in \cite{Schaye2015}.

\end{itemize}

\section{Conclusions}
\label{conclusion}
This paper introduces a public \sql relational database\footnote{Available at
  the address \dbAddress} containing the integrated quantities and merger
histories for more than $10^5$ galaxies from the \eagle suite of hydrodynamic
simulations. The database contains all the galaxies from the largest \eagle
simulation as well as galaxies from smaller volumes where the resolution and AGN
model were varied. The details of these simulations are presented by
\citet{Schaye2015} and a list of published results using the simulation can be
found on our websites\footnote{\url{http://eagle.strw.leidenuniv.nl/}
  and\\ \url{http://www.eaglesim.org}}.

For each galaxy in the database and at each redshift, we provide a wide range of
basic halo and galaxy properties such as stellar masses, gas masses, unextincted
magnitudes, angular momenta, star formation rates and $gri$ images, as well as
extensive information on metal abundances. Three additional tables give the
properties of galaxies measured in a series of apertures, more physically
motivated galaxy sizes and galaxy photometry. Using their merger trees, galaxies
can be tracked through time and their assembly history explored by analysing
their progenitors.

By making the halo and galaxy data public we hope that our simulations will be
helpful both for comparison with observational data, and as a tool for gaining
physical insight into the physics of galaxy formation. 

In Section~\ref{caveats} we presented some limitations of the simulations that
should be borne in mind when using the database. In particular, caution should
be exercised because of the finite resolution of the simulations. Over time we
intend to make additional data products available as the relevant papers are
accepted for publication. These will include, among other quantities, photometry
including dust extinction and information on the morphology of the galaxies. At
later stages, we may also release merger trees with higher time resolution, more
simulations models from \cite{Crain2015} as well as the raw particle data.

The \eagle database will hopefully be a powerful resource for the community to
explore the physics of galaxy formation, and to help interpret observational
data.

\section*{Acknowledgements}
This work would have not be possible without Lydia Heck and Peter Draper's
technical support and expertise. We are grateful to all members of the Virgo
Consortium and the \eagle collaboration who have contributed to the development
of the codes and simulations used here, as well as to the people who helped with
the analysis. We thank Jaime Salcido for his help producing figure 3, Violeta
Gonzalez-Perez, Qi Guo and Claudia Lagos for useful comments on early drafts as
well as Chris Barber, Bart Clauwens and Sean McGee for testing earlier versions
of the \eagle database.  \\ This work was supported by the Science and
Technology Facilities Council (grant number ST/F001166/1); European Research
Council (grant numbers GA 267291 ``Cosmiway'' and GA 278594
``GasAroundGalaxies'') and by the Interuniversity Attraction Poles Programme
initiated by the Belgian Science Policy Office (AP P7/08 CHARM). RAC is a Royal
Society University Research Fellow.\\ This work used the DiRAC Data Centric
system at Durham University, operated by the Institute for Computational
Cosmology on behalf of the STFC DiRAC HPC Facility (www.dirac.ac.uk). This
equipment was funded by BIS National E-infrastructure capital grant
ST/K00042X/1, STFC capital grant ST/H008519/1, and STFC DiRAC Operations grant
ST/K003267/1 and Durham University. DiRAC is part of the National
E-Infrastructure.  We acknowledge PRACE for awarding us access to the Curie
machine based in France at TGCC, CEA, Bruy\`eres-le-Ch\^atel. \\ The web site
described in this paper was based on the one build for the \emph{Millennium
  Simulation} as part of the activities of the German Astrophysical Virtual
Observatory (GAVO).

\bibliographystyle{model2-names}
\bibliography{./mybibfile}

\appendix

\onecolumn

\section{Examples of more complex queries}
\label{example_scripts}
\noindent {\bf Python - Galaxy stellar mass function}. This
example\footnote{Which can also be downloaded
here: \url{http://icc.dur.ac.uk/Eagle/Database/GSMF.py}} replicates Fig. 4
from \cite{Schaye2015} comparing the galaxy stellar mass function at $z=0.1$
({\tt SnapNum}$~=27$) in 30~pkpc apertures for three of the \eagle
simulations. The link to the database is created with the module {\sc
eagleSqlTools} available from the release website\footnote{Or directly here:
\url{http://icc.dur.ac.uk/Eagle/Database/eagleSqlTools.py}} (this module serves as an
interface to access the \eagle database). After the connection is established
(on line 9), the module can submit queries to the database. Each of the chosen
table properties (in this case we have only chosen the galaxy stellar masses)
are returned in a dictionary that can be then manipulated like any other {\sc
python} dictionary. We use the {\tt GROUP BY} \sql keyword to bin the data directly on the
server and reduce the amount of data being downloaded. The output created by this script is shown in
Fig.~\ref{fig:gsmf}.
\scriptsize
\pythonexternal{example_scripts/GSMF.py}
\normalsize
\setcounter{figure}{0}
\begin{figure}[t]
\centering\includegraphics[width=8cm,angle=0]{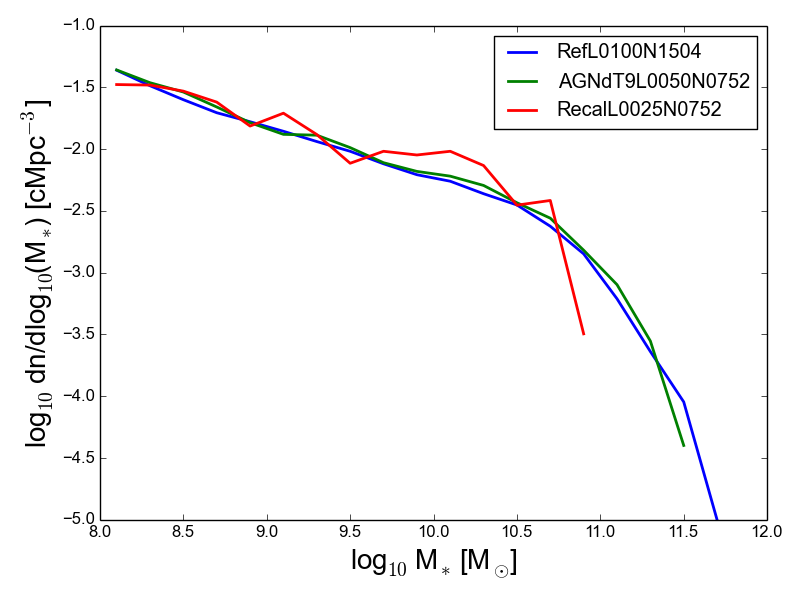}\vspace{-0.1cm}
\caption{Figure created by the {\sc python} script.}
\label{fig:gsmf}
\end{figure}
\normalsize
\FloatBarrier
\pagebreak

\noindent {\bf SQL - Black hole mass vs. stellar mass}. This example
replicates Fig. 10 from \citet{Schaye2015} showing the black hole mass as a
function of stellar mass at redshift $z=0.1$ ({\tt SnapNum}$~=27$) for 
the reference (L0100N1504) run. As mentioned in Section \ref{caveats}
we use the black hole subgrid mass, {\tt BlackHoleMass}, and treat the stellar mass of a galaxy
as the mass contained within a 30~pkpc aperture. {\bf SubHalo} table properties are connected to the {\bf
Aperture} table via each galaxy's unique \GalaxyID. 

\sqlstyle
\footnotesize
\begin{lstlisting}[numbers = none]
-- Select the quantities we want
SELECT      
	AP_Star.Mass_Star as sm,      
	SH.BlackHoleMass as bhm
-- Define aliases for the two tables      
FROM      
	RefL0100N1504_Subhalo as SH,      
	RefL0100N1504_Aperture as AP_Star
-- Apply the conditions        
WHERE      
	SH.SnapNum = 27                     -- z=0.1
	and SH.GalaxyID = AP_Star.GalaxyID  -- Match galaxies to Aperture table
	and AP_Star.ApertureSize = 30       -- Select aperture size to be 30 pkpc
	and AP_Star.Mass_Star > 0           -- Only return stellar masses > 0
	and SH.BlackHoleMass > 0            -- Only return black hole masses > 0
\end{lstlisting}

\normalsize
\noindent {\bf SQL - Galaxy size vs. stellar mass}. This example
         is similar to Fig. 9 from \citet{Schaye2015} comparing galaxy
         size as a function of stellar mass at redshift $z=0.1$ ({\tt
         SnapNum} $=27$) for each galaxy in the reference (L0100N1504)
         run. For galaxy sizes, we use the half mass radius of the
         galaxies from the {\bf Sizes} table. We
         connect them to the galaxy's stellar mass via the
         unique \GalaxyID identifier. As with the previous example, we
         must connect to the {\bf SubHalo} table to
         retrieve the {\tt SnapNum} value.

\sqlstyle
\footnotesize
\begin{lstlisting}[numbers = none]
-- Select the quantities we want
SELECT 
	AP.Mass_Star as sm, 
	SIZES.R_halfmass100 as size 
-- Define aliases for the three tables
FROM 
	RefL0100N1504_Subhalo as SH, 
	RefL0100N1504_Aperture as AP, 
	RefL0100N1504_Sizes as SIZES 
-- Apply the conditions
WHERE 
	SH.SnapNum = 27                  -- z=0.1
	and SH.GalaxyID = AP.GalaxyID    -- Match galaxies to Aperture table
	and SH.GalaxyID = SIZES.GalaxyID -- Match galaxies to Sizes table
	and AP.ApertureSize = 30         -- Select aperture size to be 30 pkpc
	and AP.Mass_Star > 0             -- Only return stellar masses > 0
\end{lstlisting}
\pagebreak

\normalsize
\noindent {\bf SQL - Joining FOF and SubHalo tables}. 
This example shows how to join the properties of galaxies to their
parent FoF halo. In this case, we compute the offset between the
centre of the potential of the galaxy and the FoF halo. When dealing 
with positions within these volumes, remember to account for box
periodicity. In principle, it is not necessary to match the {\tt
SnapNum} of the tables as well as the {\tt GroupID}, but this
speeds up the query.

\sqlstyle

\footnotesize
\begin{lstlisting}[numbers = none]
-- Select the quantities we want
SELECT              
	SH.CentreOfPotential_x as sh_x,         
	SH.CentreOfPotential_y as sh_y,         
	SH.CentreOfPotential_z as sh_z,         
	FOF.GroupCentreOfPotential_x as fof_x,         
	FOF.GroupCentreOfPotential_y as fof_y,         
	FOF.GroupCentreOfPotential_z as fof_z,
	SH.MassType_Star as mstar, 
	FOF.GroupMass as fof_mass,
	square(SH.CentreOfPotential_x-FOF.GroupCentreOfPotential_x) 
		+ square(SH.CentreOfPotential_y-FOF.GroupCentreOfPotential_y)
		+ square(SH.CentreOfPotential_z-FOF.GroupCentreOfPotential_z) as dist
-- Define aliases for the two tables
FROM  
	RefL0050N0752_Subhalo as SH,              
	RefL0050N0752_FOF as FOF
-- Apply the conditions         
WHERE  
	SH.MassType_Star > 1.0e11    -- Only return stellar masses > 1.0e11
	and SH.SnapNum = 27          -- z=0.1
	and FOF.SnapNum = SH.SnapNum -- Join SnapNum to speed up query  
	and FOF.GroupID = SH.GroupID -- Join GroupID to speed up query
\end{lstlisting}

\normalsize
\noindent {\bf SQL - Linking a progenitor to its descendants}. 
This example shows how to select a random subset of Milky Way like
galaxies, and extract information about the location and specific star
formation rates (within a 30 pkpc aperture) of all their progenitors
above a stellar mass of $10^9\Msol$.

\sqlstyle
\footnotesize
\begin{lstlisting}[numbers = none]
-- Select the quantities we want
SELECT
	DES.GalaxyID,
	PROG.Redshift,
	PROG.MassType_DM,
	PROG.MassType_Star, 
	AP.SFR / (AP.Mass_Star+0.0001) as ssfr, 
	PROG.CentreOfPotential_x,
    	PROG.CentreOfPotential_y,
	PROG.CentreOfPotential_z
-- Define aliases for the three tables                      
FROM 
	RefL0100N1504_Subhalo as PROG,      
	RefL0100N1504_Subhalo as DES,
	RefL0100N1504_Aperture as AP
-- Apply the conditions           
WHERE 
	DES.MassType_Star between 1.0e10 and 6e10     -- Select Milky Way like stellar mass
	and DES.MassType_DM between 5.0e11 and 2.0e12 -- Select Milky Way like halo mass
	and DES.RandomNumber < 0.1                    -- Take a random subset of these
	and DES.SnapNum = 28                          -- At redshift z=0            
	and PROG.GalaxyID between DES.GalaxyID and DES.LastProgID -- Then find galaxy progenitors
	and AP.ApertureSize = 30                      -- Select aperture size to be 30 pkpc            
	and AP.GalaxyID = DES.GalaxyID                -- Match galaxies to Aperture table
	and AP.Mass_Star > 1.0e9                      -- Only return galaxies with stellar mass > 1e9
-- Order the output
ORDER BY 
	DES.MassType_Star desc,
	PROG.Redshift asc,
	PROG.MassType_Star desc
\end{lstlisting}
\twocolumn

\onecolumn
\section{Description of all fields contained in the database}
\label{appendix_quantities}

\setcounter{table}{0}
\renewcommand\thetable{\Alph{section}.\arabic{table}}
\begin{table*}[h]
\caption{Full listing of the content of the main galaxy properties table and
  description of the columns. These properties are contained in tables denoted
  {\bf [modelname]\_SubHalo}. The first five lines of the table give the
  indices used to navigate between the side tables and through the merger
  trees. Particle types are dark matter, gas, stars and black holes and
  collective properties such as Mass sum over all of these particles unless otherwise stated. }
\label{table:subhalo1}
\begin{center}
\footnotesize
\renewcommand{\arraystretch}{1.5}
\begin{tabular}{ >{\ttfamily}p{4cm}p{1.5cm}p{11cm}}
{\large \bf SubHalo} & & \\
\hline
\normalfont Field & Units & Description \\
\hline\hline
GalaxyID &
- &
Unique identifier of a galaxy. This identifier enables linking the {\bf
  SubHalo} table to the {\bf Aperture}, {\bf Magnitudes} and {\bf Sizes} tables.\\

LastProgID & - & Used for merger tree traversal, Section~\ref{subsection:merger_trees}. \\

TopLeafID & - & Used for merger tree traversal, Section~\ref{subsection:merger_trees}.\\

DescendantID & - & \GalaxyID of the descendant of this galaxy, Section~\ref{subsection:merger_trees}.\\

\hline

GroupID & - & Unique identifier of the FoF halo hosting this galaxy. This identifier enables linking 
the {\bf SubHalo} table to the {\bf FOF} table.\\

\hline

Redshift & - & Redshift at which these properties are computed. \\

Snapnum & - & Snapshot number at which these properties are computed. \\

\hline

GroupNumber & - & Integer identifier of the FoF halo hosting this galaxy. GroupNumber
is only unique to a given snapshot and hence cannot be used to identify the
same halo across multiple snapshots.\\

SubGroupNumber & - & Integer identifier of this galaxy within its FoF halo. SubGroupNumber is only unique to a given FoF halo in a given snapshot and hence cannot be used to identify the same galaxy across multiple outputs. The condition ``{\tt   SubGroupNumber}$~=~0$'' selects central galaxies.\\ 

\hline

Spurious & - & Value is $1$ if the galaxy is an artefact of
the \subfind algorithm and $0$ if the galaxy is a genuine object, see Section~\ref{caveats}.\\

\hline

Image\_face &  & \\
Image\_edge & - & \\
Image\_box &  & \multirow{-3}{11cm}{Weblink to the mock $gri$
image of the galaxy in the three different orientations (face-on, edge-on and along the simulation $z$-axis). When querying the database via the browser, the image appears in the column of the results table. } \\

\hline

BlackHoleMass &
M$_\odot$ &
Sum of all the black hole subgrid masses in this galaxy. See Eq.~(\ref{eq:m_bh}) for a description of the subgrid mass $\tilde m$ of a black hole, and Section~\ref{caveats} for cautions using black hole masses. \\

BlackHoleMassAccretionRate &
M$_\odot$~yr$^{-1}$ &
Total instantaneous accretion rate of all black holes, see Section~\ref{caveats} for cautions. \\

CentreOfMass\_x & &\\
CentreOfMass\_y & cMpc &\\
CentreOfMass\_z & & \multirow{-3}{11cm}{Co-moving position of the centre of mass, Eq.~(\ref{eq:CentreOfMass}).} \\

CentreOfPotential\_x & &\\
CentreOfPotential\_y & cMpc &\\
CentreOfPotential\_z & & \multirow{-3}{11cm}{Co-moving position of the minimum of the gravitational potential defined by the position of the most bound particle.} \\

\hline
 
\end{tabular}
\end{center}
\end{table*}

\setcounter{table}{0}
\renewcommand\thetable{\Alph{section}.\arabic{table}}
\begin{table*}
\caption{ -- continued}
\label{table:subhalo2}
\begin{center}
\footnotesize
\renewcommand{\arraystretch}{1.5}
\begin{tabular}{ >{\ttfamily}p{4cm}p{1.5cm}p{11cm}}
{\large \bf SubHalo} & & \\
\hline
\normalfont Field & Units & Description \\
\hline\hline

GasSpin\_x & & \\
GasSpin\_y & pkpc~km~s$^{-1}$ & \\
GasSpin\_z & & \multirow{-3}{11cm}{Spin per unit mass of all gas particles, ${\bf L}/M$, with ${\bf L}$ given by Eq.~(\ref{eq:spin}).}\\

HalfMassRad\_DM &
pkpc &
Physical radius enclosing half of the dark matter mass.\\

HalfMassRad\_Gas &
pkpc &
Physical radius enclosing half of the gas mass.\\

HalfMassRad\_Star &
pkpc &
Physical radius enclosing half of the stellar mass.\\

HalfMassRad\_BH &
pkpc &
Physical radius enclosing half of the black hole particle mass, $m$, as defined in Eq.~(\ref{eq:m_bh}). \\

HalfMassProjRad\_DM &
pkpc &
Projected physical radius enclosing half of the dark matter mass, averaged over three orthogonal projections.\\

HalfMassProjRad\_Gas &
pkpc &
Projected physical radius enclosing half of the gas mass, averaged over three orthogonal projections.\\

HalfMassProjRad\_Star &
pkpc &
Projected physical radius enclosing half of the stellar mass, averaged over three orthogonal projections.\\

HalfMassProjRad\_BH &
pkpc &
Projected physical radius enclosing half of the black hole particle mass, $m$ (Eq.~\ref{eq:m_bh}), averaged over three orthogonal projections.\\

InitialMassWeightedBirthZ &
z &
Mean redshift of formation of stars, weighted by birth mass $\tilde m$ (Eq.\ref{eq:m_star}). Calculated via $\sum_i \tilde {m}_{i} \tilde z_i / \sum_i \tilde {m}_{i}$ where $\tilde z_i$ is the redshift
the star particle $i$ was formed and $\tilde m_i$ its birth mass.\\

InitialMassWeightedStellarAge &
Gyr &
Mean age of stars, weighted by birth mass. Calculated via $\sum_i \tilde {m}_{i} (t - \tilde t_i) / \sum_i \tilde {m}_{i}$ where $t$ is cosmic time, and $\tilde t_i$ and $\tilde m_i$ formation time and birth mass of the star particle $i$, respectively.\\

KineticEnergy &
M$_{\odot}$~(km/s)$^{2}$ &
Total kinetic energy $E_K$, Eq.(\ref{eq:KineticEnergy}).\\

Mass &
M$_\odot$ &
Total current mass of all particles (i.e. $\sum_i m_i$ where $m_{i}$ is the mass of the particle).\\

MassType\_DM &
M$_\odot$ &
Total dark matter mass. \\

MassType\_Gas &
M$_\odot$ &
Total gas mass. \\

MassType\_Star &
M$_\odot$ &
Total stellar mass, $\sum_i m_i$, where $m_i$ is the stellar particle mass from Eq.~(\ref{eq:m_star}).\\

MassType\_BH &
M$_\odot$ &
Total black hole mass, $\sum_i m_i$, where $m_i$ is the black hole particle mass from Eq.~(\ref{eq:m_bh}).\\

RandomNumber &
- &
Random number uniform in the range $[0,1)$. \\

StarFormationRate &
M$_\odot$~yr$^{-1}$ &
Total star formation rate, $\sum_i \dot m_{\star, i}$, where $\dot m_{\star, i}$ is the star formation rate of gas particle $i$. \\

StellarInitialMass &
M$_\odot$ &
Sum of birth masses of all stars, $\sum_i \tilde{m}_{i}$, where $\tilde {m}_{i}$ is the birth mass of star particle $i$ from Eq.~(\ref{eq:m_star}).\\

StellarVelDisp &
km~s$^{-1}$ &
One dimensional velocity dispersion of stars, $((2\,\texttt{Star\_KineticEnergy})/(3\,\texttt{MassType\_Star}))^{1/2}$, where
${\tt Star\_KineticEnergy}$ is the kinetic energy of stars.\\

ThermalEnergy &
M$_\odot$~(km/s)$^{2}$ &
Total thermal energy $E_u$, Eq.~(\ref{eq:ThermalEnergy}).\\

TotalEnergy &
M$_\odot$~(km/s)$^{2}$ & 
Total energy $E_{\rm tot}$, Eq.~(\ref{eq:TotalEnergy}).\\

Velocity\_x & & \\
Velocity\_y & km~s$^{-1}$ & \\
Velocity\_z & & \multirow{-3}{11cm}{Peculiar velocity, Eq.~(\ref{eq:Velocity}).}\\

Vmax &
km~s$^{-1}$ & Maximum value of the circular velocity, $(G(M(<r)/r))^{1/2}$, where $M(<r)$ is the total mass enclosed in a sphere of physical radius $r$.\\

VmaxRadius &
pkpc &
Physical radius where the circular velocity equals \texttt{Vmax}.\\

\hline

\end{tabular}
\end{center}
\end{table*}

\setcounter{table}{0}
\renewcommand\thetable{\Alph{section}.\arabic{table}}
\begin{table*}
\caption{ -- continued. Columns in this table exist for each of three different components:
star-forming gas ({\tt SF}), non-star-forming gas ({\tt NSF}) and
stars ({\tt Stars}). As these properties are repeated for each of these components,
we only describe them once. In the database each property will be preceded with
either {\tt [SF/NSF/Stars]\_} before its name. For instance, the metallicity
field will exist in three variants: {\tt SF\_Metallicity}, {\tt NSF\_Metallicity}
and {\tt Stars\_Metallicity} for the metallicity of the star-forming gas, of the
non star-forming gas and of the stars, respectively. Any sum used to describe a property is the sum of all particles for that component only. }
\label{table:subhalo3}
\begin{center}
\footnotesize
\renewcommand{\arraystretch}{1.5}
\begin{tabular}{ >{\ttfamily}p{4cm}p{1.5cm}p{11cm}}
{\large \bf SubHalo} & & \\
\hline
\normalfont Field & Units & Description \\
\hline\hline

Hydrogen & - & \\
Helium & - & \\
Carbon & - & \\
Nitrogen & - & \\
Oxygen & - & \\
Neon & - & \\
Magnesium & - & \\
Silicon & - & \\
Sulphur & - & \\
Calcium & - & \\
Iron & - & \multirow{-11}{11cm}{Total mass in this element divided by the total mass (both for a given component). These are therefore absolute abundances which do not depend on the solar abundance.} \\

IronFromSNIa &
- &
Total mass in {\tt Iron} contributed by ejecta from Type Ia supernovae, divided by the total mass.\\

KineticEnergy &
M$_{\odot}$~(km/s)$^{2}$ &
Total kinetic energy $E_K$, Eq.~(\ref{eq:KineticEnergy}).\\

Mass &
M$_{\odot}$ &
Total mass, $\sum_i m_i$, where $m_i$ is the particle mass.\\

MassFromAGB &
M$_{\odot}$ &
Total mass contributed by ejecta of AGB stars.\\

MassFromSNII &
M$_{\odot}$ & 
Total mass contributed by ejecta from massive stars and type II supernovae.\\

MassFromSNIa &
M$_{\odot}$ & 
Total mass contributed by ejecta from type Ia SN supernovae. \\

MassWeightedEntropy &
km~s$^{2}$~$(\frac{10^{10} \Msol}{\rm Mpc})^\frac{2}{3}$ &
~~~~~~Mass-weighted pseudo entropy of all particles, $\sum_i m_i S_i / \sum_i m_i$, where $S_{i}$ is the pseudo-entropy (Eq.~\ref{eq:S}) and $m_{i}$ is the mass of the particle $i$. (Entry present for gas components only.)\\

MassWeightedTemperature &
K &
Mass-weighted temperature, $\sum_i m_i \mathrm{T}_{i} / \sum_i m_{i}$, where $\mathrm{T}_{i}$ is the temperature and $m_{i}$ is the mass of the particle $i$. (Entry present for gas components only.)\\

Metallicity &
- & 
Metal mass fraction, $\sum_i m_i Z_i / \sum_i m_i$, where $Z_{i}$ is the metallicity and $m_{i}$ is the mass of the particle $i$. \\

MetalsFromAGB &
M$_\odot$ & 
Total metal mass contributed by ejecta from AGB stars. \\

MetalsFromSNII &
M$_\odot$ & 
Total metal mass contributed by ejecta from massive stars and SN Type II supernovae.\\

MetalsFromSNIa &
M$_\odot$ & 
Total metal mass contributed by ejecta from Type Ia supernovae.\\

Spin\_x & & \\
Spin\_y & pkpc~km~s$^{-1}$ &  \\
Spin\_z & & \multirow{-3}{11cm}{Spin per unit mass, ${\bf L}/M$, from Eq.~(\ref{eq:spin}).}\\

ThermalEnergy & 
M$_\odot$~(km/s)$^{2}$ &
Total thermal energy $E_u$, Eq.~(\ref{eq:ThermalEnergy}). (Entry present for gas components only.)\\

TotalEnergy & 
M$_\odot$~(km/s)$^{2}$ &
Total energy $E_{\rm tot}$, Eq.~(\ref{eq:TotalEnergy}). The potential energy contribution does include the other components as well, $E_{\Phi}={1\over 2}\sum_i m_i{\hat\Phi_i\over a}$.\\

\hline 

\end{tabular}
\end{center}
\end{table*}

\begin{table*}
\caption{Full listing of the content of the halo table and description of
  the columns. These properties are contained in tables denoted {\bf
    [modelname]\_FOF}. This table can be linked to the {\bf
    [modelname]\_SubHalo} table using the unique {\tt GroupID}
    identifier.}
\label{table:fof}
\begin{center}
\footnotesize
\renewcommand{\arraystretch}{1.5}
\begin{tabular}{ >{\ttfamily}p{4cm}p{1.5cm}p{11cm}}
{\large \bf FOF} & & \\
\hline
\normalfont Field & Units & Description \\
\hline\hline

GroupID &
- &
Unique identifier of a halo (i.e. a FoF group). This identifier enables linking a halo
to all its galaxies and their properties in the {\bf SubHalo} table.\\

\hline

Redshift & - & Redshift at which these properties are computed. \\

SnapNum & - & Snapshot number containing that halo. \\

\hline

GroupCentreOfPotential\_x &
 & \\
GroupCentreOfPotential\_y &
cMpc & \\
GroupCentreOfPotential\_z & & \multirow{-3}*{Co-moving position of the minimum of the gravitational potential of the halo.}\\

GroupMass &
M$_{\odot}$ &
Total Friends-of-Friends mass of this halo. \\

Group\_M\_Crit200  \\

Group\_M\_Crit500  &  M$_{\odot}$ \\

Group\_M\_Crit2500 & & \multirow{-3}{11cm}{Total mass within the corresponding {\tt Group\_R\_Critxxx} radius, where xxx=200, 500 or 2500, respectively} \\

Group\_M\_Mean200 \\ 
Group\_M\_Mean500 & M$_{\odot}$ \\
Group\_M\_Mean2500 & & \multirow{-3}{11cm}{Total mass within the corresponding {\tt Group\_R\_Meanxxx} radius, where xxx=200, 500 or 2500, respectively} \\

Group\_M\_TopHat200 &
M$_{\odot}$ &
Total mass within radius {\tt Group\_R\_Tophat200}. \\

Group\_R\_Crit200 \\
Group\_R\_Crit500 & pkpc \\
Group\_R\_Crit2500 & & \multirow{-3}{11cm}{Physical radius within which the mean density is xxx times the \emph{critical} density of the Universe, where xxx=200, 500 or 2500, respectively}.\\

Group\_R\_Mean200\\
Group\_R\_Mean500 & pkpc \\
Group\_R\_Mean2500 & & \multirow{-3}{11cm}{Physical radius within which the mean density is xxx times the \emph{mean} density of the Universe, where xxx=200, 500 or 2500, respectively}.\\

Group\_R\_TopHat200 &
pkpc &
Physical radius within which the mean density is $18\pi^{2} + 82
(\Omega_{m}(z)-1)-39 (\Omega_{m}(z)-1)^{2}$ times the critical density of the Universe. This is based on 
the spherical top-hat collapse model of \citep{Bryan1998}.\\

NumOfSubhalos &
- &
Number of subhaloes (galaxies) identified as belonging to this halo. \\

RandomNumber &
- &
Random number uniform in the range $[0,1)$. \\
\hline
\end{tabular}
\end{center}
\end{table*}

\begin{table*}
\caption{Full listing of the content of the galaxy sizes table and description of
  the columns. These properties are contained in tables denoted {\bf
    [modelname]\_Sizes}. This table contains half-mass sizes of
    the {\em stellar} component of galaxies using spherical apertures (Furlong et al. (\textit{in prep.}). The \GalaxyID column can be
    used to join this table to the corresponding {\bf
    [modelname]\_SubHalo} table. Only galaxies with total stellar mass $M_*>10^8\Msol$
    have entries in this table. }
\label{table:sizes}
\begin{center}
\footnotesize
\renewcommand{\arraystretch}{1.5}
\begin{tabular}{ >{\ttfamily}p{4cm}p{1.5cm}p{11cm}}
{\large \bf Sizes} & & \\
\hline
\normalfont Field & Units & Description \\
\hline\hline

GalaxyID &
- &
Unique identifier of a galaxy as per {\bf SubHalo} table. \\

\hline

R\_halfmass30 &
pkpc &
Half mass radius of stellar component within a spherical (3D) 30~pkpc aperture.\\

R\_halfmass100 &
pkpc &
Half mass radius of stellar component within a spherical (3D) 100 pkpc aperture.\\

R\_halfmass30\_projected &
pkpc &
Projected half mass radius of stellar component within a circular (2D) 30~pkpc aperture
(averaged over three orthogonal projections).\\

R\_halfmass100\_projected & pkpc & Projected half mass radius of stellar component within a circular
(2D) 100~pkpc aperture (averaged over three orthogonal
projections).\\
\hline

\end{tabular}
\end{center}
\end{table*}

\begin{table*}
\caption{Full listing of the content of the aperture table and description of
  the columns. These properties are contained in tables denoted {\bf
    [modelname]\_Aperture}. This table contains measurements within spherical
  apertures centred on the minimum of the gravitational potential of a given galaxy. Each row
  represents a set of measurements for a single galaxy using a single aperture
  size in physical kpc. The \GalaxyID column can be used to join this table to
  the corresponding {\bf [modelname]\_SubHalo} table.}
\label{table:aperture}

\begin{center}
\footnotesize

\renewcommand{\arraystretch}{1.5}
\begin{tabular}{ >{\ttfamily}p{4cm}p{1.5cm}p{11cm}}
{\large \bf Aperture} & & \\
\hline
\normalfont Field & Units & Description \\
\hline\hline

GalaxyID &
- &
Unique identifier of a galaxy as per {\bf SubHalo} table.\\

\hline

ApertureSize &
pkpc &	
Spherical (3D) aperture radius used for this measurement. Quantities are measure in a sphere centred at the centre of the potential, {\em i.e.} at the location of the most bound particle. Available aperture sizes are: 1, 3, 5, 10, 20, 30, 40, 50, 70 and 100~pkpc. \\

VelDisp &
km~s$^{-1}$ &
One dimensional velocity dispersion of stars, $((2\,\textrm{KineticEnergy\_Star})/(3\,\textrm{Mass\_Star}))^{1/2}$, where ${\rm KineticEnergy\_Star}$ is the kinetic energy of stars, and the sum is over stars within the aperture.\\

SFR &
M$_{\odot}$~yr$^{-1}$ &
Star formation rate within the aperture. \\

Mass\_BH &
M$_{\odot}$ &	
Total particle mass, $\sum_i m_i$ (Eq.~\ref{eq:m_bh}), of all black holes within the aperture. \\

Mass\_DM &
M$_{\odot}$ &	
Total dark matter mass within the aperture. \\

Mass\_Gas &
M$_{\odot}$ &	
Total gas mass within the aperture. \\

Mass\_Star &
M$_{\odot}$ &	
Total stellar mass, $\sum_i m_i$ (Eq.~\ref{eq:m_star}), within the aperture. \\

\hline
\end{tabular}
\end{center}
\end{table*}

\begin{table*}
\caption{Full listing of the content of the magnitudes table and description of
  the columns. These properties are contained in tables denoted {\bf
    [modelname]\_Magnitudes}. This table contains absolute rest-frame magnitudes without dust
    attenuation for all galaxies with $M_*>10^{8.5}\Msol$ contained in the {\bf
    SubHalo} table. This table can be joined to the {\bf SubHalo} table using
    the \GalaxyID field. The magnitudes in the different
    SDSS \citep{SDSSfilters} and UKIRT \citep{UKIRTfilters} filters have been
    computed in 30~pkpc spherical apertures following the procedure described
    by \citet{Trayford2015}.}
\label{table:magnitudes}
\begin{center}
\footnotesize
\renewcommand{\arraystretch}{1.5}
\begin{tabular}{ >{\ttfamily}p{4cm}p{1.5cm}p{11cm}}
{\large \bf Magnitudes} & & \\
\hline
\normalfont Field & Units & Description \\
\hline\hline

GalaxyID &
- &
Unique identifier of a galaxy as per {\bf SubHalo} table.\\

\hline
u\_nodust & mag & Rest-frame absolute magnitude (AB) in the $u$ band without dust
attenuation. \\
g\_nodust & mag & Rest-frame absolute magnitude (AB) in the $g$ band without dust
attenuation. \\
r\_nodust & mag & Rest-frame absolute magnitude (AB) in the $r$ band without dust
attenuation. \\
i\_nodust & mag & Rest-frame absolute magnitude (AB) in the $i$ band without dust
attenuation. \\ 
z\_nodust & mag & Rest-frame absolute magnitude (AB) in the $z$ band without dust
attenuation. \\ 
Y\_nodust & mag & Rest-frame absolute magnitude (AB) in the $Y$ band without dust
attenuation. \\ 
J\_nodust & mag & Rest-frame absolute magnitude (AB) in the $J$ band without dust
attenuation. \\ 
H\_nodust & mag & Rest-frame absolute magnitude (AB) in the $H$ band without dust
attenuation. \\ 
K\_nodust & mag & Rest-frame absolute magnitude (AB) in the $K$ band without dust
attenuation. \\ 
\hline

\end{tabular}
\end{center}
\end{table*}

\twocolumn

\section{List of snapshot output times} 
\label{appendix_snapshot}
\setcounter{table}{0}
\begin{table}[h]
\caption{List of all output redshifts for all the simulations present in the
  database. Note that \sql queries are made easier by the use of the snapshot number
  rather than the redshift. Lookback times are given in Gigayears.}
\label{table:snapshot_list}
\begin{center}
\footnotesize
\renewcommand{\arraystretch}{1.5}
\begin{tabular}{rrrr}
\hline
{\tt SnapNum} & {\tt Redshift} & Lookback  time & Expansion factor \\
\hline\hline
28 &  0.00 & 0.00  & 1.000  \\ 
27 &  0.10 & 1.34  & 0.909  \\
26 &  0.18 & 2.29  & 0.846  \\
25 &  0.27 & 3.23  & 0.787  \\
24 &  0.37 & 4.16  & 0.732  \\
23 &  0.50 & 5.19  & 0.665  \\
22 &  0.62 & 6.01  & 0.619  \\
21 &  0.74 & 6.71  & 0.576  \\
20 &  0.87 & 7.37  & 0.536  \\
19 &  1.00 & 7.93  & 0.499  \\
18 &  1.26 & 8.86  & 0.443  \\
17 &  1.49 & 9.49  & 0.402  \\
16 &  1.74 & 10.05 & 0.365  \\
15 &  2.01 & 10.53 & 0.332  \\
14 &  2.24 & 10.86 & 0.309  \\
13 &  2.48 & 11.16 & 0.287  \\
12 &  3.02 & 11.66 & 0.249  \\
11 &  3.53 & 12.01 & 0.221  \\
10 &  3.98 & 12.25 & 0.201  \\
9  &  4.49 & 12.46 & 0.182  \\
8  &  5.04 & 12.63 & 0.166  \\
7  &  5.49 & 12.75 & 0.154  \\
6  &  5.97 & 12.86 & 0.143  \\
5  &  7.05 & 13.04 & 0.124  \\
4  &  8.07 & 13.16 & 0.110  \\
3  &  8.99 & 13.25 & 0.100  \\
2  &  9.99 & 13.32 & 0.091  \\
1  &  15.13& 13.53 & 0.062  \\
0  &  20.00& 13.62 & 0.047  \\

\hline
\end{tabular}

\end{center}
\end{table}

\normalsize
\section{Detailed expressions for quantities in the database}
\label{appendix_cosmo}
In order to remove any ambiguity in the quantities provided in the database, this appendix summarises the fundamental equations that are being solved and the co-ordinate system used in the numerical code. 

The equations that describe the evolution of a gravitating fluid are the continuity, Euler, energy and Poisson equations \citep{Peebles1980}. In order to provide a precise definition of the symbols used to describe database entries, we write these equations as 
\begin{eqnarray}
\frac{\partial\rho}{\partial t} + ({\bf v}\cdot \nabla)\,\rho \equiv {{\rm d}\rho\over {\rm d}t}&=&-\rho\nabla\cdot {\bf v}\label{eq:cont}\\
{{\rm d}{\bf v}\over {\rm d}t}&=&-\frac{1}{\rho}\nabla p-\nabla\Phi\label{eq:euler}\\
{{\rm d}u\over {\rm d}t}&=& -{p\over \rho}\nabla\cdot{\bf v}
-{\cal C}\rho\,\,\\
\nabla^2\Phi  &=&4\pi G (\rho+\rho_{\rm col}) - \Lambda\,,
\label{eq:CEP}
\end{eqnarray}
where $\rho$ is gas density, $\rho_{\rm col}$ the density due to collisionless matter (i.e., stars, dark matter and black holes), $p$ the effective gas pressure, $\Phi$ the (Newtonian) gravitational potential and $\Lambda$ the cosmological constant. The variable
\begin{equation}
u = {p\over (\gamma-1)\rho} = {k_{\rm B}T\over (\gamma-1)\,\mu m_{\rm H}}\,, 
\label{eq:u}
\end{equation}
is the thermal energy per unit mass, with $\gamma=5/3$ the ratio of specific heats for a mono-atomic gas,
and $\mu$ the mean molecular weight in units of the Hydrogen mass, $m_{\rm H}$. The term ${\cal C}(T)\,\rho$ describes radiative cooling and heating.
The database also reports the value of the pseudo-entropy $S$, defined as
\begin{equation}
S \equiv {p\over\rho^\gamma}\,.
\label{eq:S}
\end{equation}

We use the standard notation for proper time, $t$, position, ${\bf r}$, and velocity ${\bf v}\equiv {\rm d}{\bf r}/{\rm d}t\equiv \dot{\bf r}$. Partial derivatives are defined so that $\partial/\partial t$ is the time derivative at constant position ${\bf r}$,
${\rm d}/{\rm d}t\equiv \partial/\partial t + ({\bf v}\cdot\nabla)$ is the Lagrangian time derivative, and the spatial derivative $\nabla\equiv \partial/\partial_{\bf r}$ is computed at constant time.

We solve these equations in an expanding coordinates described by the scale factor $a(t)$ which satisfies the Friedmann equations,
\begin{eqnarray}
H^2 \equiv \left({{\dot a}\over a}\right)^2 &=& {8\pi {\rm G}\over 3}\bar\rho_t+ {\Lambda\over 3}\\
\ddot a &=& -{4\pi {\rm G}\over 3}\bar\rho_t\,a + {\Lambda\over 3}a\,,
\end{eqnarray}
where $\Lambda\equiv 3H_0^2\Omega_\Lambda$ (with $H_0\equiv H(a=1)$), and $\bar\rho_t$ is the mean total density, $\bar\rho_t = \bar\rho + \bar\rho_{\rm col}$. We apply periodic boundary conditions in this expanding reference frame.

The simulation uses comoving coordinates to simplify the integration of Eqs.~\ref{eq:cont}-\ref{eq:CEP}. These are defined as
\begin{eqnarray}
{\bf x}&\equiv & {{\bf r}\over a}\\
\hat \rho &\equiv & a^3\,\rho\\
\hat u &\equiv& a^{-2}u\\
\hat\Phi &=& a\,(\Phi -{2\pi\over 3}{\rm G}\bar\rho_t r^2 + {1\over 6}\Lambda\,r^2)\\
\hat\nabla &\equiv &a\,\nabla\\
\hat p &=& (\gamma-1)\hat \rho\,\hat u\\
\hat S &=& S\,.
\label{eq:covar}
\end{eqnarray}
In these variables, the velocity
\begin{eqnarray}
{\bf v} &=& \dot a {\bf x} + {\bf v}_p\\
{\bf v}_p & \equiv & a \dot{\bf x}\,,
\label{eq:v_p}
\end{eqnarray}
where ${\bf v}_p$ is referred to as the peculiar velocity. We will use the term \lq comoving variable\rq\ when a quantity is expressed in comoving variables (i.e. ${\bf x}$, $\dot{\bf x}$ and hatted variables), and \lq physical\rq\ or \lq proper\rq\ otherwise. In particular we will express comoving distances in cMpc (for comoving mega parsecs) and physical distances in pMpc (for proper or physical mega parsecs), and similarly for ckpc and pkpc.

The equations are solved by representing the collisionless mass as well as the gas by particles. We denote particle masses as $m_i$, for particle $i$. In {\sc Eagle}\,, star particles lose mass to gas particles to represent mass loss from stars. Each star particle therefore has two mass variables, its current particle mass, $m$, and its birth mass $\tilde m$:
\begin{eqnarray}
m &=& \hbox{current particle mass of star}\nonumber\\
\tilde m&=&\hbox{birth mass of star}\,,
\label{eq:m_star}
\end{eqnarray}
see \cite{Wiersma2009b} for more details. Black holes also have two mass variables associated with them, a particle mass $m$, and a subgrid mass $\tilde m$. It is the subgrid mass that enters the equations describing the accretion rate of a black hole. In short,
\begin{eqnarray}
m &=& \hbox{particle mass of black hole}\nonumber\\
\tilde m&=&\hbox{subgrid mass of black hole}\,.
\label{eq:m_bh}
\end{eqnarray}
Once a black hole is significantly more massive than the seed mass, particle and subgrid mass trace each other closely, see \cite{Booth_Schaye2009} and \cite{Guevara2013} for details.

Having defined comoving variables, the comoving energy $\hat E$ of a collisionless halo is
\begin{equation}
\hat E = {1\over 2}\sum_i m_i (a^2\dot{\bf x}_i)^2 + {1\over 2} \sum_i {m_i\,\hat\Phi_i}\,,
\end{equation}
and is conserved for an isolated halo, as is its comoving spin $\hat{\bf L}$,
\begin{equation}
\hat{\bf L} = \sum_i m_i({\bf x}-{\bf x}_{\rm com})\times(a^2\dot{\bf x}_i-a^2{\dot{\bf x}}_{\rm com})\,.
\label{eq:spin}
\end{equation}
Here 
\begin{equation}
{\bf x}_{\rm com}=\sum_i m_i{\bf x}_i/\sum_i m_i\,,
\label{eq:com}
\end{equation}
is the comoving position of the centre of mass (taking into account periodic boundary conditions), and $\dot{\bf x}_{\rm com}$ its time derivative.

The database uses comoving co-ordinates to record the locations of haloes and galaxies.
For example, the position of the centre of a galaxy or halo (stored as {\tt CentreOfMass} in the database) is 
\begin{equation}
\hbox{\footnotesize CentreOfMass} = {\bf x}_{\rm com} = {\sum_i m_i\,{\bf x}_i\over \sum_i m_i}\,,
\label{eq:CentreOfMass}
\end{equation}
where the sum runs over all particles that belong to the object taking into account periodic boundary conditions. Similarly, the centre of potential of an object (database variable {\tt CentreOfPotential}) is given in comoving coordinates.

The velocity of a halo or galaxy (database variable {\tt Velocity}) refers to its peculiar velocity,
\begin{equation}
\hbox{\footnotesize Velocity} = a\,\dot{\bf x}_{\rm com} = a{\sum_i m_i\,\dot{\bf x}_i\over \sum_i m_i}\,.
\label{eq:Velocity}
\end{equation}

All other variables are expressed in physical coordinates, for example the spin vector of a galaxy is computed as
\begin{equation}
\hbox{\footnotesize Spin} = {\sum_i m_i\,({\bf r}_i-{\bf r}_{\rm com})\times ({\bf v}_i-{\bf v}_{\rm com})\over \sum_i m_i}\,,
\end{equation}
where ${\bf r}_{\rm com}$ and ${\bf v}_{\rm com}$ are the physical position and velocity of the centre of mass.
The expressions for physical kinetic, potential, thermal, and total energy are, respectively
\begin{eqnarray}
E_{\rm K} &=&{1\over 2}\sum_i\,m_i\,({\bf v}-{\bf v}_{\rm com})^2 \label{eq:KineticEnergy}\\
E_{\Phi} &=&{1\over 2}\sum_i m_i{\hat\Phi_i\over a} \label{eq:PotentialEnergy}\\
E_{\rm u} &=& \sum_i\,m_i\,(a^2\hat u_i) \label{eq:ThermalEnergy}\\
E_{\rm tot} &=& E_{\rm K}+E_{\Phi}+E_{\rm u}\,. \label{eq:TotalEnergy}
\end{eqnarray}

\end{document}